\RequirePackage{etex} 
\documentclass[runningheads]{llncs}

\usepackage{color}
\usepackage[final]{graphicx}
\usepackage{algorithmic}
\usepackage{epsfig}
\usepackage{amssymb}
\usepackage{amsmath}
\usepackage{multicol}
\usepackage{colortbl}
\usepackage{booktabs}
\usepackage{multirow}
\usepackage[subtle]{savetrees}

\usepackage[utf8]{inputenc}

\usepackage{amsfonts}
\usepackage{graphicx}
\usepackage[table,xcdraw]{xcolor}
\graphicspath{ {images/} }
\usepackage{crypto/crypto}
\usepackage{crypto/protocol}
\usepackage[boxruled,vlined,linesnumbered]{algorithm2e}
\usepackage[n ,advantage ,operators ,sets, adversary,landau,probability,notions,logic,ff,mm,primitives,events,complexity,asymptotics,keys]{cryptocode}
\usepackage{caption}
\usepackage{subcaption}
\usepackage{adjustbox}

\usepackage{float}
\usetikzlibrary{matrix,shapes,arrows,positioning,chains, calc}

\usepackage{tikz}
\usetikzlibrary{fit}
\usetikzlibrary{shapes,arrows,positioning}
\newlength{\nodedistance}

\usepackage{amsthm} 
\theoremstyle{plain}

\usepackage{hyperref}
\usepackage{centernot}

\usepackage{lscape}

\usepackage{enumitem}
\usepackage{bigdelim}
\usepackage{adjustbox}
\usepackage{subcaption}
\usepackage{bm}
\usepackage{orcidlink}

\usepackage[]{todonotes}

\SetKwData{Left}{left}\SetKwData{This}{this}\SetKwData{Up}{up}
\SetKwFunction{Union}{Union}\SetKwFunction{FindCompress}{FindCompress}
\SetKwInOut{Input}{Input}\SetKwInOut{Output}{Output}

\newcommand{\seeda}{\ensuremath{\text{seed}_{\pmb{A}}}}
 
\newcommand{\KEM}{\ensuremath{\mathtt{KEM}} }

\usepackage{parskip}

\begin{document}

\title{On the Masking-Friendly Designs for Post-Quantum Cryptography}

\author{Suparna Kundu\inst{1} \and Angshuman Karmakar\inst{1,2} \and Ingrid Verbauwhede\inst{1}}

\authorrunning{Kundu et al.}

\institute{COSIC, KU Leuven, Kasteelpark Arenberg 10, Bus 2452, B-3001 Leuven-Heverlee, Belgium \and
  Indian Institute of Technology Kanpur, India
  \\\email{{firstname.lastname}@esat.kuleuven.be}}

\maketitle
 
\begin{abstract}

Masking is a well-known and provably secure countermeasure against side-channel attacks. However, due to additional redundant computations, integrating masking schemes is expensive in terms of performance. The performance overhead of integrating masking countermeasures is heavily influenced by the design choices of a cryptographic algorithm and is often not considered during the design phase. 

In this work, we deliberate on the effect of design choices on integrating masking techniques into lattice-based cryptography. We select Scabbard, a suite of three lattice-based post-quantum key-encapsulation mechanisms (KEM), namely Florete, Espada, and Sable. We provide arbitrary-order masked implementations of all the constituent KEMs of the Scabbard suite by exploiting their specific design elements. We show that the masked implementations of Florete, Espada, and Sable outperform the masked implementations of Kyber in terms of speed for any order masking. Masked Florete exhibits a $73\%$, $71\%$, and $70\%$ performance improvement over masked Kyber corresponding to the first-, second-, and third-order. 
Similarly, Espada exhibits $56\%$, $59\%$, and $60\%$ and Sable exhibits $75\%$, $74\%$, and $73\%$ enhanced performance for first-, second-, and third-order masking compared to Kyber respectively. 
Our results show that the design decisions have a significant impact on the efficiency of integrating masking countermeasures into lattice-based cryptography.

\end{abstract}

\keywords{Post-quantum cryptography \and Key-encapsulation mechanism \and Side-channel attacks \and Scabbard \and Higher-order masking}

\section{Introduction}\label{sec:intro}

Physical attacks such as fault injection and side-channel attacks are potent threats to any cryptosystem deployed in the public domain. Classical cryptographic schemes such as elliptic-curve cryptography~\cite{ECC_miller_Crypto86} and RSA~\cite{RSA} went through decades of testing, analysis, and invention of different physical attacks and their countermeasures to generate enough confidence to be successfully deployed in the real world. In comparison, post-quantum cryptography (PQC), or specifically lattice-based cryptography (LBC) has gone through significantly less amount of investigation in the context of physical attacks. Therefore, although the United States government's National Institute of Standards and Technology (NIST) has recently proposed some standard PQC schemes~\cite{nist_final_report}, for a successful transition to PQC, it is imperative that we concentrate our research efforts in this direction.

Masking~\cite{ChaJutRaoRoh1999} is an interesting countermeasure against passive physical attacks or side-channel attacks (SCA) such as power analysis, electromagnetic radiation analysis, etc. On a fundamental level, masking works by splitting the secret into multiple random shares and performing the same computation as the unmasked version on each share. Thus, the security of masking is based on the same information-theoretic principles, such as Shamir's secret sharing~\cite{shamir_secret_sharing} or multi-party computation~\cite{mpc_andrew}. Masking can provide provably secure countermeasures against side-channel attacks. 
Nevertheless, due to the duplication of computations, the runtime of a masked implementation theoretically grows significantly with the increase in the order of masking.
For example, in the case of Kyber, a post-quantum key-encapsulation mechanism (KEM) scheme that has been selected as standard in the NIST's procedure, the runtime of the first, second, and third order of masked implementation is 12, 20, and 30 times of the unmasked implementation on ARM Cortex-M4 platform \cite{BronchainC22}.

Our primary motivation in this work is to assess how the design decisions of a lattice-based KEM scheme, such as the choice of quotient polynomial, distribution of secrets and errors, underlying hard problems, modulus, etc., influence their masking performance. We also want to test how close we can get to the theoretical upper bound of efficiency in masking.  For our experiments, we have chosen the post-quantum KEM suite Scabbard~\cite{Bermudo_Mera_Karmakar_Kundu_Verbauwhede_2021} with 3 different lattice-based schemes. First, a ring-learning with rounding (RLWR) based scheme Florete with ring size comparable to NewHope~\cite{DBLP:conf/uss/AlkimDPS16}, second a module-learning with rounding (MLWR) based scheme Sable with ring size similar to Saber~\cite{Saber_kem} and Kyber~\cite{Kyber-Kem}, and finally an MLWR-based scheme Espada with unique smaller ring size. The choice of Scabbard helps us to demonstrate our methods on diverse KEM schemes with many variations in the design. Scabbard was proposed to improve the NIST PQC finalist KEM Saber~\cite{Saber_kem}. The designers of Scabbard argued that all the design decisions of Scabbard had been propelled by the experience gained in the research and developments in the field of lattice-based cryptography of previous years.
Therefore, it inherits all the advantages of Saber \textit{i.e.} less randomness due to rounding, power-of-two modulus for efficient masking, simple algorithms for efficiency and faster deployment on diverse platforms, etc. Further, the design of Scabbard improves in areas like suitability for parallel implementation, flexibility, efficiency, and adaptation of faster masking schemes. We will discuss the schemes of Scabbard in Sec.~\ref{sec:prelims_scabbard}. In the original publication~\cite{Bermudo_Mera_Karmakar_Kundu_Verbauwhede_2021}, the authors have provided different implementations on hardware and software platforms to prove their claims on efficiency.  
It was shown before that the design of Saber is highly conducive to masking~\cite{FO-masked-saber}. 
Due to these reasons, Scabbard is an ideal choice to demonstrate the interplay between design choices and masking performance in lattice-based KEMs.

In this work, we propose arbitrary-order masked implementations of all the KEMs in the suite Scabbard. We implement and benchmark them on an ARM Cortex-M4 microcontroller platform using the PQM4~\cite{PQM4} library to prove the masking friendliness of its design. The ring size of the polynomial length matches the number of message bits, which is $256$ for Saber or Kyber as well as Sable. So, the encoding of message bits to the ciphertext polynomial is trivial in these cases. 
However, this is not the case for Florete and Espada, and these schemes use $\mathtt{original\_msg}$ function for message decoding and $\mathtt{arrange\_msg}$ function for message encoding. This work introduces a higher-order masked version of $\mathtt{original\_msg}$ and $\mathtt{arrange\_msg}$ function. These functions can be applied to all LWR-based KEMs with different ring sizes than 256 and even learning with errors (LWE) based KEMs with some modifications. The schemes of Scabbard use different centered binomial distributions compared to Saber or Kyber. For this purpose, we modified the masked centered binomial distribution (CBD) algorithms proposed by Schneider et al. \cite{Tobias2019_binomial_sampler} for each scheme of Scabbard and optimized it for them. Public and re-encrypted ciphertext comparison is an important part of the Fujisaki-Okamoto transformation used in LWE-/LWR-based KEM. It is faster for unmasked or first-order masking but becomes computationally expensive for higher-order maskings. Here, we modified the ciphertext comparator of \cite{HO_mask_Saber} for each scheme of Scabbard to obtain better performance. These masked components are faster in Scabbard than Kyber, thanks to the choice of RLWR/ MLWR hard problem, power-of-two moduli and slightly reduced parameter sets.

As performance results, the overhead factor we obtained for masked Florete for the first-, second-, and third-order are approximately $2.7$x, $5$x, and $7.7$x, compared to the unmasked implementation. For Espada, the overhead cost of the first-, second-, and third-order masked versions are roughly $1.8$x, $2.8$x, and $4$x than the unmasked one. The performance cost of masked Sable for the first-, second-, and third-order are around $2.4$x, $4.3$x, and $6.3$x over the unmasked version. We compare the masked implementations of Florete, Espada, and Sable with the state-of-the-art masked implementation of Kyber and Saber. We show that the masked implementations of all the schemes of Scabbard surpass the masked implementations of Kyber in terms of performance for any order masking, and masked implementations of Florete and Sable outperform masked implementations of Saber for arbitrary order. More specifically, masked Florete performs $73\%$, $71\%$, and $70\%$ better than masked Kyber, corresponding to the first-, second-, and third-order. Espada shows  $56\%$, $59\%$, and $60\%$ performance improvement for first-, second- and third-order masked implementations compared to Kyber. Masked Sable exceeds the execution time of masked Kyber by $75\%$, $74\%$, and $73\%$ for the first-, second-, and third-order. Our masked implementations are available at \url{https://github.com/Suparna-Kundu/Masked_Scabbard.git}.
%\url{https://anonymous.4open.science/r/Masked_Scabbard-698A}.

To conclude this section, we want to draw attention to the fact that although the NIST standardization procedure for PKE/KEM has been finalized with Kyber, we firmly believe that further investigations and innovations are required to improve side-channel secure PQC schemes. 
The NIST procedure opened the possibility of exploring different possibilities to improve various aspects of PQC schemes. We have witnessed this throughout the course and even after the NIST procedure. For example, Mitaka~\cite{mitaka} has been proposed, which is a masking-friendly version of Falcon~\cite{web:falcon}, a NIST standard for digital signatures. Kyber-90s version of Kyber was proposed to use the advanced encryption standard (AES) as a pseudo-random number generator instead of the slower Keccak extended output function. Similarly, Saber-90s and uSaber were proposed as alternate versions of the NIST PQC standardization finalist scheme Saber to improve efficiency and ease of masking. As discussed earlier, Scabbard~\cite{Bermudo_Mera_Karmakar_Kundu_Verbauwhede_2021} was an improvement of Saber. The design of Scabbard has further influenced the design of PQC KEM Smaug~\cite{smaug_kem}, which is a candidate scheme from ongoing Korean PQC standardization \cite{kpqc}. Therefore, exploring various design choices and their effect on different aspects of the performance of existing PQC schemes is an interesting research direction.

\section{Preliminaries}\label{sec:prelims}

% \subsection{Notations}

For a positive integer $q$, the set of integers modulo $q$ is denoted by $\mathbb{Z}_q$. The quotient ring $\mathbb{Z}_q[x]/f(x)$ is denoted by $\mathcal{R}_q^n$, where $f(x)$ is a $n$ degree cyclotomic polynomial over $\mathbb{Z}_q[x]$. We use lowercase letters to denote an element of this ring, which is a polynomial. We indicate the ring of $l$ length vectors over the ring $\mathcal{R}_q^n$ as $(\mathcal{R}_q^n)^l$ and use bold lowercase letters to denote an element of this ring which is a vector of polynomials. The ring of $l \times l$ length matrices over the ring $\mathcal{R}_q^n$ as $(\mathcal{R}_q^n)^{l \times l}$. The elements of this ring are $l \times l$ matrices of polynomials and are represented by uppercase letters. $x \leftarrow \mathcal{\chi} (S)$ represents that $x$ is sampled from the set $S$ and follows the distribution $\chi$. When $x$ is generated using a pseudo-random number generator expanding a seed $seed_x$ over the set $S$, we denote it as $x \leftarrow \chi(S; seed_x)$.
%When the $S$ is generated using a pseudo-random number generator expanding a seed $seed_x$ we denote it as $x \leftarrow \chi(S; seed_x)$.
%If an element $x$ is constructed from a seed $seed_x$ with some pseudorandom number $\chi$, then it is denoted by $v \leftarrow \chi(S; seed_v)$. 
We use $\mathcal{U}$ to denote the uniform distribution and the CBD whose standard deviation $\sqrt{\mu/2}$ is presented by $\mathcal{\beta_\mu}$. We denote the rounding operator with $\lfloor \cdot \rceil$, which returns the closest integer and is rounded upwards during ties. These operations can be extended over the polynomials by applying them coefficient-wise. The polynomial multiplication between two polynomials of length $n$ is represented using $n \times n$ multiplication. We use $\{x_i\}_{0 \leq i \leq t}$ to represent the set $\{ x_0,\ x_1,\ \ldots,\ x_t \}$ which contains $t+1$ elements of the ring $\mathcal{R}$.

\subsection{Scabbard: a Post-Quantum KEM Suite}\label{sec:prelims_scabbard}

Scabbard is a suite of post-quantum KEMs proposed by Mera et al. \cite{Bermudo_Mera_Karmakar_Kundu_Verbauwhede_2021} that improved state-of-the-art LBC schemes by incorporating different design choices and newer developments in the field. The security of the schemes in the Scabbard depends on some variants of learning with rounding (LWR) problems, more specifically, module-LWR (MLWR) and ring-LWR (RLWR) problems. Banerjee et al.~\cite{Banerjee_lwr} introduced the LWR problem and also showed that the LWR problem is as hard as the LWE problem. If ${A} \leftarrow \mathcal{U}((\mathbb{Z}_q)^{l\times l})$, secret $\mathbf{s} \leftarrow \beta_\mu((\mathbb{Z}_q)^{l})$, error $\mathbf{e} \leftarrow \beta_{\mu_e}((\mathbb{Z}_q)^{l})$, and $\mathbf{b} \leftarrow \mathcal{U}((\mathbb{Z}_q)^{l})$ then distinguishing between $(A,\ As + e)$ and $(A,\ b)$ is hard and this problem is known as the decision version of LWE problem. The decision version of the LWR problem states that if ${A} \leftarrow \mathcal{U}((\mathbb{Z}_q)^{l\times l})$, secret $\mathbf{s} \leftarrow \beta_\mu((\mathbb{Z}_q)^{l})$, and for some $p<q$, $\mathbf{b} \leftarrow \mathcal{U}((\mathbb{Z}_p)^{l})$ then distinguishing between $(A,\ \lfloor (q/p) As \rceil)$ and $(A,\ b)$ is hard \cite{Banerjee_lwr}. In the LWR problem, the explicit sampling of error $\mathbf{e}$ in the LWE is replaced by the rounding operation. %In case of the MLWR problem, ${A} \leftarrow \mathcal{U}((\mathcal{R}_q^n)^{l\times l})$, $\mathbf{s} \leftarrow \beta_\mu((\mathcal{R}_q^n)^{l})$, $\mathbf{b} \leftarrow \mathcal{U}((\mathcal{R}_p^n)^{l})$ and the problem is similar to the MLWE problem, $(A,\ \lfloor (q/p) As \rceil)$ and $(A,\ b)$ are statistically indistinguishable \cite{Langlois2015}. 
In case of the MLWR problem, ${A} \leftarrow \mathcal{U}((\mathcal{R}_q^n)^{l\times l})$, $\mathbf{s} \leftarrow \beta_\mu((\mathcal{R}_q^n)^{l})$, $\mathbf{b} \leftarrow \mathcal{U}((\mathcal{R}_p^n)^{l})$ and the MLWR problem states that $(A,\ \lfloor (q/p) As \rceil)$ and $(A,\ b)$ are computationally indistinguishable \cite{Langlois2015}. 
%Considering $l=1$ in the MLWR problem, we can receive the RLWR problem \cite{ringlwe_2010}. So, in the description of generic LWR-based KEM for Scabbard, the MLWR problem is used. 
%In standard LWR-based and RLWR-based constructions, the matrix size of the underlying problem are $l$ and $n$, respectively. On the other hand, MLWR-based construction works as tradeoffs between standard LWR-based and RLWR-based structures. Therefore, the matrix size of the underlying problem is $l \times n$ in MLWR-based constructions. It makes the structures of MLWR-based construction more generic, as we can convert the MLWR-based scheme to a standard LWR-based scheme by fixing $n = 1$ and an RLWR-based scheme by setting $l = 1$. So, the MLWR-based construction is used to describe a generic LWR-based KEM in Scabbard.
In standard LWR-based and RLWR-based constructions, the ranks of underlying matrices are respectively $l$ and $n$, with very high probability. On the other hand, MLWR-based constructions are proposed as a trade-off between standard LWR-based and RLWR-based structures. The rank of underlying matrices in MLWR-based schemes is $l \times n$. It makes the structures of MLWR-based constructions more generic, as we can convert the MLWR-based scheme to a standard LWR-based one by fixing $n = 1$ and an RLWR-based one by setting $l = 1$. %So, the MLWR-based construction is used to describe a generic LWR-based KEM in Scabbard.
Therefore, we use MLWR notations to describe the schemes in Scabbard below.
A KEM needs to be secure against chosen ciphertext attacks (IND-CCA/IND-CCA2: indistinguishable against a-posteriori chosen-ciphertext attacks). In LWR-based KEM, it is accomplished by applying Jiang \textit{et al.}'s version \cite{Jiang_2017} of Fujisaki-Okamoto (FO) transformation \cite{FujisakiO99} over the generic LWR-based public-key encryption (PKE), where the PKE needs to be secure against chosen plaintext attacks (IND-CPA: indistinguishable against chosen plaintext attack). We denote generic LWR-based PKE as \texttt{LWR.PKE} and generic LWR-based KEM as \texttt{LWR.KEM}, which are shown respectively in \autoref{fig:saberpke} and \autoref{fig:saberkem}. In \texttt{LWR.KEM}, $\mathcal{H}$, $\mathcal{G}$, and $\mathtt{KDF}$ three hash functions are required as part of FO transformation. This suite of KEMs consists of three schemes: (i) Florete, (ii) Espada, and (iii) Sable. We briefly describe these three schemes with their specific features below.

\vspace{-4mm}
\begin{figure}[ht]
% \begin{adjustbox}
\fbox{\begin{varwidth}{\textwidth}
\begin{subfigure}[t]{0.49\textwidth}
    \raggedright 
    \begin{small}
    $\mathtt{LWR}{.}\mathtt{PKE}{.}\mathtt{KeyGen} ()$
    \begin{enumerate}[wide=0em, itemsep=0pt, parsep=0pt, font=\scriptsize\tt\color{gray}]
            \item $seed_{\pmb{A}} \leftarrow  \mathcal{U}(\{0,1\}^{256})$ \\
            \item $\pmb{A} \leftarrow \mathcal{U}({\text{(}\mathcal{R}_q^n\text{)}}^{l\times l};\ \seeda) $ \\
            \item $r \leftarrow \mathcal{U}(\{0,\ 1\}^{256})$ \\
            \item $\pmb{s} \leftarrow \beta_\mu ({\text{(}\mathcal{R}_q^n\text{)}}^l;\ r)$ \\
            \item $\pmb{b} =  ((\pmb{A}^T \pmb{s} + \pmb{h}) \bmod{q}) \gg (\epsilon_q - \epsilon_p) \in {\text{(}\mathcal{R}_p^n\text{)}}^l$ \\
            \item \textbf{return} $(pk = (seed_{\pmb{A}},\ \pmb{b}),\ sk =  (\pmb{s}))$
    \end{enumerate} 
    \end{small}
\end{subfigure}
\begin{subfigure}[t]{0.50\textwidth}
    \raggedright
    \begin{small}
    $\mathtt{LWR}{.}\mathtt{PKE}{.}\mathtt{Enc} (pk = (seed_{\pmb{A}}, \pmb{b}), m \in R_2; r )$
    \begin{enumerate}[wide=0em, itemsep=0pt, parsep=0pt, font=\scriptsize\tt\color{gray}]
        \item $\pmb{A} \leftarrow \mathcal{U}({\text{(}\mathcal{R}_q^n\text{)}}^{l\times l};\ \seeda) $ \\
        \item \textbf{if: } $r$  is not specified:
        \item \quad $ r \leftarrow \mathcal{U}(\{0,\ 1\}^{256})$
        \item $\pmb{s'} \leftarrow \beta_\mu ( {\text{(}\mathcal{R}_q^n\text{)}}^l ;\ r)	$
        \item $\pmb{u} = ( ( \pmb{A}  \pmb{s}' + \pmb{h}) \bmod{q})  \gg (\epsilon_q - \epsilon_p)  \in {\text{(}\mathcal{R}_p^n\text{)}}^l$
        \item $ c_m = \pmb{b}^T  (\pmb{s}' \bmod{p})  \in \mathcal{R}_p^n$
        \item $v = (c_m  + h_1 - 2^{\epsilon_p-B} m \bmod{p}) \gg (\epsilon_p - \epsilon_t - B) \in \mathcal{R}_{2^{B}t}^n $
        \item \textbf{return} $c =(\pmb{u},\ v)$
    \end{enumerate} 
    \end{small}
\end{subfigure}
\begin{subfigure}[t]{\textwidth}
   \vspace{-.8cm}
    \raggedright
    \begin{small}
    $\mathtt{LWR}{.}\mathtt{PKE}{.}\mathtt{Dec}(sk=\pmb{s},c=(\pmb{u}, v))$
    \begin{enumerate}[wide=0em, itemsep=0pt, parsep=0pt, font=\scriptsize\tt\color{gray}]
        \item $u'' = \pmb{u}^{T} (\pmb{s} \bmod{p}) \in \mathcal{R}_p^n $\\
        \item $m'' = (u''-2^{\epsilon_p-\epsilon_t-B}v + h_2) \bmod{p}$
        \item $m' = m'' \gg (\epsilon_p-B) \in \mathcal{R}_{2^{B}t}^n$  \\
        \item \textbf{return} $m'$
    \end{enumerate} 
    \end{small}
\end{subfigure}
\end{varwidth}}
\caption{Generic $\mathtt{LWR}{.}\mathtt{PKE}$~\cite{Bermudo_Mera_Karmakar_Kundu_Verbauwhede_2021}}
\label{fig:saberpke}
% \end{adjustbox}
\end{figure}
\vspace{-6mm}
\begin{figure}[ht]
\fbox{\begin{varwidth}{\textwidth}
\begin{subfigure}[t]{0.47\textwidth}
    \begin{small}
    \raggedright 
    $\mathtt{LWR}{.}\mathtt{KEM}{.}\mathtt{KeyGen} ()$
    \begin{enumerate}[wide=0em, itemsep=0pt, parsep=0pt, font=\scriptsize\tt\color{gray}]
        \item $(seed_{\pmb{A}}, \pmb{b}, \pmb{s})  = \mathtt{LWR}{.}\mathtt{PKE}{.}\mathtt{KeyGen} ()$ \\
        \item $pk = (seed_{\pmb{A}}, \pmb{b})$ \\
        \item $pkh = \mathcal{H}(pk) $ \\
        \item $z \leftarrow \mathcal{U}(\{0,\ 1\}^{256}$) \\
        \item \textbf{return} $(pk = (seed_{\pmb{A}},\ \pmb{b}),\ sk =  (\pmb{s},\ z,\ pkh))$
    \end{enumerate} 
    \end{small}
    \vspace{3mm}
\end{subfigure}
\begin{subfigure}[t]{0.47\textwidth}
    \begin{small}
    \raggedright
    $\mathtt{LWR}{.}\mathtt{KEM}{.}\mathtt{Encaps} (pk = (seed_{\pmb{A}},\ \pmb{b}))$
    \begin{enumerate}[wide=0em, itemsep=0pt, parsep=0pt, font=\scriptsize\tt\color{gray}]
        \item $m^\prime  \leftarrow  \mathcal{U}(\{0,1\}^{256}) $ \\
        \item $m = \mathtt{arrange\_msg}(m^\prime)$
        \item $m  = \mathcal{H}(m)$ \\
        \item $(\hat{K}, r) = \mathcal{G}(\mathcal{H}(pk), m)$ \\
        \item $c = \mathtt{LWR}{.}\mathtt{PKE}{.}\mathtt{Enc} (pk, m; r )$ \\
        \item $K = \mathtt{KDF}(\hat{K}, \mathcal{H}(c))$
        \item \textbf{return} $(c, K)$
    \end{enumerate} 
    \end{small}
\end{subfigure}
\begin{subfigure}[t]{\textwidth}
    \begin{small}
    \raggedright
    $\mathtt{LWR}{.}\mathtt{KEM}{.}\mathtt{Decaps} (sk = (\pmb{s},\ z,\ pkh),pk = (seed_{\pmb{A}},\ \pmb{b}),\ c)$
    \begin{enumerate}[wide=0em, itemsep=0pt, parsep=0pt, font=\scriptsize\tt\color{gray}]
        \item $m''  =  \mathtt{LWR}{.}\mathtt{PKE}{.}\mathtt{Dec} (\pmb{s}, c )$ \\
        \item $m'= \mathtt{original\_msg}(m''$)\\
        \item $(\hat{K}', r') = \mathcal{G}(pkh, m')$ \\
        \item $c_* = \mathtt{LWR}{.}\mathtt{PKE}{.}\mathtt{Enc} (pk, m'; r')$ \\
        \item \textbf{if: } $c=c_*$
        \item \quad \textbf{return} $ K = \mathtt{KDF}(\hat{K}', \mathcal{H}(c))$
        \item \textbf{else: }
        \item \quad \textbf{return} $ K = \mathtt{KDF}(z, \mathcal{H}(c))$  
    \end{enumerate} 
    \end{small}
\end{subfigure}
\end{varwidth}}
\caption{Generic $\mathtt{LWR}{.}\KEM$~\cite{Bermudo_Mera_Karmakar_Kundu_Verbauwhede_2021}}
\label{fig:saberkem}
% \vspace{-10pt}
\end{figure}
\vspace{-5mm}

\subsubsection{Florete:}\label{florete}

%This scheme is designed to achieve faster running time. It is based on the RLWR problem, and so the value of the parameter $l$ in Figure \ref{fig:saberpke} is always $1$. 
This scheme is based on the RLWR problem \textit{i.e.} $l=1$ in Figure.\ref{fig:saberpke} and designed for faster running time.
Here, the cyclotomic polynomial used to construct the quotient rings $\mathcal{R}_q^n$, $\mathcal{R}_p^n$, and $\mathcal{R}_t^n$ is $(x^{768}-x^{384}+1)$. In Florete, one message bit is encoded in three coefficients of the polynomial $v$ in line $7$ of \texttt{LWR.PKE.Enc} algorithm of Figure \ref{fig:saberpke}. So, during the encapsulation process, as shown in line $2$ of \texttt{LWR.KEM.Encaps} algorithm of Figure \ref{fig:saberkem}, a conversion from $256$ bits of message to a polynomial of length $768$ is performed with the help of $\mathtt{arrange\_msg}$ function and it is defined as: $\mathtt{arrange\_msg}(m') = m'||m'||m'\,.$
The inverse of $\mathtt{arrange\_msg}$ function is used in the \texttt{LWR.KEM.Decaps} algorithm named as $\mathtt{original\_msg}$, and the $\mathtt{original\_msg}: \mathbb{Z}_2^{768} \longrightarrow \mathbb{Z}_2^{256} $ is defined as if $\mathtt{original\_msg}(m'') = m'$ and $b\in \{ 0,\ 1,\ \ldots,\ 255 \}$ then $
m'[b] =
\begin{cases}
0 & \text{if } m''[b]+m''[b+256]+m''[b+512]\leq 1 \\
1 & \text{otherwise } \\
\end{cases}\,.$ 
In Florete, $768 \times 768$ polynomial multiplication is used, and it is performed using the combination of Toom-Cook $3$-way, Toom-Cook $4$-way, $2$ levels of Karatsuba, and $16 \times 16$ schoolbook multiplication.
\vspace{-10pt}
\subsubsection{Espada:}\label{espada}

This scheme is designed to reduce the memory footprint on software platforms. It is based on the MLWR problem, and the cyclotomic polynomial is used to construct the underlying quotient ring of the lattice problem $\mathcal{R}_q^n$ is $(x^{64}+1)$. The polynomial length here is 64, so the dimension of vectors of polynomial $l$ is taken equal to $12$ to maintain security. In Espada, the $256$ bit message is encoded inside the $64$ length polynomial $v$, so four message bits are encoded in a coefficient of the polynomial $v$. The $\mathtt{arrange\_msg}: \mathbb{Z}_2^{256} \longrightarrow \mathbb{Z}_4^{64}$ and the function is defined as: $\mathtt{arrange\_msg}(m') = m''$, where for $b \in \{0,\ 1,\ \ldots,\ 63\}$
\begin{equation}
    m''[b] = m'[4*b+3]||m'[4*b+2]||m'[4*b+1]||m'[4*b]. \label{eq1}
\end{equation}

The $\mathtt{original\_msg}: \mathbb{Z}_4^{64} \longrightarrow \mathbb{Z}_2^{256}$ function is defined as: $\mathtt{original\_msg}(m'') = m'$ and follows Equation \ref{eq1}. Lastly, the $64 \times 64$ polynomial multiplication is performed using $2$ levels of Karatsuba and $16 \times 16$ schoolbook multiplication.
\vspace{-10pt}
\subsubsection{Sable:}\label{improved_saber}
This scheme can be interpreted as an alternate version of Saber and is designed to improve performance with less memory footprint. It is also based on the MLWR problem, and similar to Saber, the cyclotomic polynomial used here in the quotient rings is $(x^{256}+1)$. The $\mathtt{arrange\_msg}$ function and $\mathtt{original\_msg}$ function are described as: $\mathtt{arrange\_msg}(m') = m'$ and $\mathtt{original\_msg}(m'') = m'' = m'$, respectively. The polynomial multiplication used in Sable is identical to Saber. The $256 \times 256$ polynomial multiplication is realized by the combination of Toom-Cook $4$-way, $2$ levels of Karatsuba, and $16 \times 16$ schoolbook multiplication.

The concrete security of these schemes depends on the parameter set, which includes the three power-of-two ring moduli $t<p<q$, the length of a polynomial $n$, the dimension of the vector of polynomial $l$, the CBD parameter $\mu$, and the number of message-bit encoded in a coefficient of the polynomial is represented by $B$. Table \ref{table:parameters} presents the parameter sets for all three schemes that achieve the NIST security level 3. We humbly refer to the original Scabbard paper \cite{Bermudo_Mera_Karmakar_Kundu_Verbauwhede_2021} for more insightful details.

% Please add the following required packages to your document preamble:
% \usepackage[table,xcdraw]{xcolor}
% If you use beamer only pass "xcolor=table" option, i.e. \documentclass[xcolor=table]{beamer}
\begin{table}[!ht]
\vspace{-6mm}
\centering
\scriptsize
\caption{Parameters of Scabbard suite}
\label{table:parameters}
\begin{tabular}{ccccccccccc}
\hline
{ Scheme Name} & \multicolumn{2}{c}{{ \begin{tabular}[c]{@{}c@{}}Ring/Module\\ Parameters\end{tabular}}} & { \begin{tabular}[c]{@{}c@{}}PQ\\ Security\end{tabular}} & { \begin{tabular}[c]{@{}c@{}}Failure\\ probability\end{tabular}} & \multicolumn{2}{c}{{ Moduli}} & { \begin{tabular}[c]{@{}c@{}}CBD\\ ($\beta_\eta$)\end{tabular}} & { Encoding} & \multicolumn{2}{c}{{ \begin{tabular}[c]{@{}c@{}}Key sizes for \\ KEM (Bytes)\end{tabular}}} \\ \hline
{ }            & {n:}                                                 & {768}                                                & { }                                                      & { }                                                              & {$\epsilon_q$:}               & {10}              &                                                                                     &                                 & { {Public key:}}                                  & {896}                                   \\
{ {Florete}}   &                                                      &                                                      & { {$2^{157}$}}                                           & { {$2^{-131}$}}                                                  & {$\epsilon_p$:}               & {9}               & {$\eta=1$}                                                                          & {B=1}                           & { {Secret key:}}                                  & {1152}                                  \\
{ }            & {l:}                                                 & {1}                                                  & { }                                                      & { }                                                              & {$\epsilon_t$:}               & {3}               &                                                                                     &                                 & { {Ciphertext:}}                                  & {1248}                                  \\ \hline
{ }            & {n:}                                                 & {64}                                                 & { }                                                      & { }                                                              & {$\epsilon_q$:}               & {15}              &                                                                                     &                                 & { {Public key:}}                                  & {1280}                                  \\
{ {Espada}}    &                                                      &                                                      & { {$2^{128}$}}                                           & { {$2^{-167}$}}                                                  & {$\epsilon_p$:}               & {13}              & {$\eta=3$}                                                                          & {B=4}                           & { {Secret key:}}                                  & {1728}                                  \\
{ }            & {l:}                                                 & {12}                                                 & { }                                                      & { }                                                              & {$\epsilon_t$:}               & {3}               &                                                                                     &                                 & { {Ciphertext:}}                                  & {1304}                                  \\ \hline
{ }            & {n:}                                                 & {256}                                                & { }                                                      & { }                                                              & {$\epsilon_q$:}               & {11}              &                                                                                     &                                 & { {Public key:}}                                  & {896}                                   \\
{ {Sable}}     &                                                      &                                                      & { {$2^{169}$}}                                           & { {$2^{-143}$}}                                                  & {$\epsilon_p$:}               & {9}               & {$\eta=1$}                                                                          & {B=1}                           & { {Secret key:}}                                  & {1152}                                  \\
{ }            & {l:}                                                 & {3}                                                  & { }                                                      & { }                                                              & {$\epsilon_t$:}               & {4}               &                                                                                     &                                 & { {Ciphertext:}}                                  & {1024}                                  \\ \hline
\end{tabular}
\vspace{-4mm}
% \raggedbottom
\end{table}

\subsection{Masking}\label{masking}

%Many recent works have shown that lattice-based public-key cryptographic schemes are vulnerable to side-channel attacks (SCA)\cite{ACL+2020,GJN2020,RBS+2020,catinca_sca}. Masking \cite{ChaJutRaoRoh1999,DBLP:conf/ches/ChariRR02} is a well-developed, provably secure countermeasure against SCA. 
The effectiveness of masking against SCA has been well demonstrated for symmetric-key block ciphers %\cite{DBLP:conf/ches/RivainP10,DBLP:conf/ches/KimHL11,DBLP:conf/eurocrypt/Coron14} 
\cite{DBLP:conf/ches/RivainP10,DBLP:conf/eurocrypt/Coron14} 
and recently extended for LBC~\cite{FO-masked-saber,HO_mask_Saber,ho-mask-comparator-kyber-bos-2021}. In $n$-th order masking, we split the sensitive data $x$ into $(n+1)$ shares and perform all the operations on each share separately. So, an adversary with a limited number of probes, such as at most $n$ probes, does not receive any advantages compared to another adversary who does not have access to those probes. The $n$th order masking technique can prevent up to $n$th order differential power attacks. However, the integration of masking techniques in LBC schemes affects the performance of the algorithm significantly with the increment of the masking order. The design decision of cryptographic schemes affects the performance of masked versions of the lattice-based schemes. This is why even though the unmasked performance of NIST finalist Saber is almost the same as Kyber, the masked version of Saber is way faster than masked Kyber for any masking order. Masked version Saber gains this advantage thanks to the choice of LWR problem and power-of-two moduli. The KEMs in the suite Scabbard also use power-of-two moduli and further improve the efficiency of the LWR-based schemes. In this work, we investigate whether the efficiency of Scabbard will translate to the masked domain.

\section{Masking Scabbard}\label{sec:masking_scabbard}

\begin{figure}
     \centering
\setlength{\nodedistance}{4mm}
\resizebox{.9\textwidth}{!}{%
\begin{tikzpicture}[ % <--- used style are moved here 
node distance = \nodedistance,
operation/.style = {circle, draw=black},
bigoperation/.style = {rounded corners, draw=black},
operationsecret/.style = {circle, draw=black, fill=black!20!white, text=black, very thick},
bigoperationsecret/.style = {rounded corners, draw=black, fill=black!20!white, text=black, very thick},
operant/.style = {},
halfblockdraw/.style = {draw, rounded corners},
     line/.style = {draw, -Latex}
                       ]
% Place nodes
% bp * s
\node (multbps) [operationsecret]  {X};
\node (bp) [operant, left=of multbps.west]  {$\pmb{u}$};
\draw[->] (bp) -- (multbps);

\node (s) [operant, above=of multbps.north]  {$\pmb{s}$};
\draw[->, very thick] (s) -- (multbps);

% v + h addition
\node (addvh) [operationsecret, right=of multbps]  {$+$};
\draw[->, very thick] (multbps.east) -- (addvh.west);
\node (h2) [operant, above=of addvh.north]  {$\pmb{h}_2$};
\draw[->] (h2) -- (addvh);

% cm
\node (cm) [operant, below=of bp]  {$v$};
\node (shiftcm) [bigoperation, right=of cm]  {$\ll$};
\draw[->] (cm) -- (shiftcm);

% cm - v
\coordinate [right=of addvh](addvhright);
\node (subcmv) [operationsecret, very thick] at (addvhright |- shiftcm) {$-$};
\draw[->, very thick] (addvh) -- (addvhright) -- (subcmv);
\draw[->] (shiftcm) -- (subcmv);

% tobin0
\node (tobin0) [bigoperationsecret, right=of subcmv]  {$\gg$};
\draw[->, very thick] (subcmv) -- (tobin0);

% % tobin1
% \node (tobin1) [bigoperationsecret, right=of tobin0]  {$\mathtt{OMsg}$};
% \draw[->, very thick] (tobin0) -- (tobin1);

% tobin
\node (tobin) [bigoperationsecret, right=of tobin0]  {$\mathtt{OMsg}$};
\draw[->, very thick] (tobin0) -- (tobin);

% G(pk, m)
\node (g) [bigoperationsecret, right=of tobin]  {$\mathcal{G}$};
\coordinate  (m) at ($(tobin.east)!0.5!(g.west)$);
\node (pkh) [operant, above=of g]  {$pkh$};
\draw[->] (pkh) -- (g);
\draw[->, very thick] (tobin) -- (g);

\node (K) [operant, below=of g]  {$\hat{K}'$};
\draw[->, very thick] (g) -- (K);

% gen secret
\node (beta1) [bigoperationsecret, right=of g]  {XOF};
\draw[->, very thick] (g) -- (beta1);

\node (beta2) [bigoperationsecret, right=of beta1]  {$\beta_\mu$};
\draw[->, very thick] (beta1) -- (beta2);

% a * sp
\node (multasp) [operationsecret, right=of beta2]  {X};
\draw[->, very thick] (beta2) -- (multasp);

\node (gena) [bigoperation, above=of multasp]  {$\mathcal{U}$};
\draw[->] (gena) -- (multasp);

\node (seeda) [operant, above=of gena]  {$\seeda$};
\draw[->] (seeda) -- (gena);

% b * sp
\node (multbsp) [operationsecret, below=of multasp]  {X};
\coordinate  (r) at ($(beta2.east)!0.4!(multasp.west)$);
\draw[->, very thick]  (r) -- (r |- multbsp) -- (multbsp); 

\node (b) [operant, below=of multbsp]  {$\pmb{b}$};
\draw[->] (b) -- (multbsp);

% add h1
\node (addh1) [operationsecret, right=of multasp]  {$+$};
\draw[->, very thick] (multasp) -- (addh1);
\node (h1) [operant, above=of addh1]  {$\pmb{h}$};
\draw[->] (h1) -- (addh1);

% add h
\node (addh2) [operationsecret, right=of multbsp]  {$+$};
\draw[->, very thick] (multbsp) -- (addh2);
\node (h) [operant, below=of addh2]  {$h_1$};
\draw[->] (h) -- (addh2);

%b2a
\node (b2a) [bigoperationsecret, below= 4.5\nodedistance of beta1)]  {$\mathtt{AMsg}$};

% add m
\node (addm) [operationsecret, right=of addh2]  {$+$};
\coordinate [below=0.5\nodedistance of h](tmp); 
\draw[->, very thick] (addh2) -- (addm);
\draw[->, very thick] (m) -- (m |- b2a.west) -- (b2a.west);
\draw[->, very thick] (b2a.east) -- (addm |- b2a.east) -- (addm);

% % add m
% \node (addm) [operationsecret, right=of addh2]  {$+$};
% \coordinate [below=0.5\nodedistance of h](tmp); 
% \draw[->, very thick] (addh2) -- (addm);
% \draw[->, very thick] (m) -- (m |- tmp) -- (addm |- tmp) -- (addm);

% divide
\node (divide2) [bigoperationsecret, right=of addm]  {$\gg$};
\node (divide1) [bigoperationsecret] at (divide2 |- addh1)  {$\gg$};
\draw[->, very thick] (addm) -- (divide2);
\draw[->, very thick] (addh1) -- (divide1);

% b*, v*
\node (bp2) [operant, right=of divide1]  {$\pmb{u}^{*}$};
\node (cm2) [operant, right=of divide2]  {${v}^{*}$};
\draw[->, very thick] (divide1) -- (bp2);
\draw[->, very thick] (divide2) -- (cm2);

% compare
\node[draw=black, dotted, thick, fit=(bp2) (cm2)](ciphertext2) {};
\node[draw=black, dotted, thick, fit=(bp) (cm)](ciphertext) {};
\coordinate [below=1.5\nodedistance of tmp](x); 
\coordinate (y) at ($(ciphertext)!0.5!(ciphertext2)$);
\node (comp) [operationsecret, very thick] at (y |- x)  {$=$};

\draw[->] (ciphertext) -- (ciphertext |- x) -- (comp);
\draw[->, very thick] (ciphertext2) -- (ciphertext2 |- x) -- (comp);

% two end results
\node (correct) [bigoperation, below left=2\nodedistance and 2\nodedistance of comp]  {return $\mathcal{H}(\hat{K}', c)$};
\draw[->] (comp.south) -- (correct);
\node[operant,  below left=0.5\nodedistance and 1.3\nodedistance of comp] (yes) {yes};

\node (correct) [bigoperation, below right=2\nodedistance and 2\nodedistance of comp]  {return $\mathcal{H}(z, c)$};
\draw[->] (comp.south) -- (correct);
\node[operant,  below right=0.5\nodedistance and 1.3\nodedistance of comp] (no) {no};
\end{tikzpicture}}

\caption{Decapsulation of LWR-based KEM. The operations in color gray are involved with the long-term secret $\pmb{s}$ and are susceptible to side-channel attacks}
\label{fig:saberdecaps}
% \vspace{-10pt}
\end{figure}
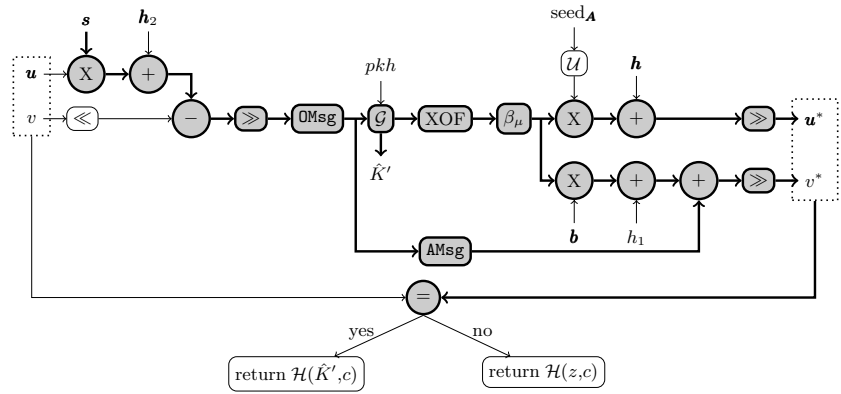

The CCA-secure KEM schemes are used to share secrets among communicating parties. Here, the secret key is non-ephemeral \textit{i.e.} the key generation is run once to generate a long-term secret key that can be used for multiple sessions and communicating with multiple entities. Therefore, in a KEM scheme, only the decapsulation is executed multiple times to retrieve the secret data from multiple entities through multiple sessions. However, this is also advantageous for an adversary. The adversary can run the decapsulation operation multiple times to improve the precision of its fault injection or take multiple side-channel traces to reduce noise in its measurements, thus improving its success probability. Mounting attacks on other operations, such as key generation and encapsulation, are relatively harder. Once an adversary compromises the secret key, it can use it to expose the secret keys of multiple sessions. Therefore, protecting the decapsulation operation from side-channel attacks is critical for the side-channel security of a KEM. We display the flow of the decapsulation algorithm of generic LWR-based KEM in Figure \ref{fig:saberdecaps} and denoted vulnerable operations in the color gray. Here $\mathtt{original\_msg}$ and $\mathtt{arrange\_msg}$ functions are shown by $\mathtt{OMsg}$ and $\mathtt{AMsg}$. In this section, we will describe the masking methods of all the components susceptible to SCA in the decapsulation operation of the Scabbard schemes.

Here, we have used two masking techniques: (i) arithmetic masking and (ii) Boolean masking to mask the Scabbard suite's schemes because these schemes consist of some operations that are cheaper to mask using arithmetic masking and some are easy to mask using Boolean masking. In both the $t$-order arithmetic and Boolean masking techniques, first we split the sensitive operand $x \in \mathbb{Z}_q = \mathbb{Z}_{2^{\epsilon_q}} = \mathbb{Z}_{2}^{\epsilon_q}$ into $(t+1)$ shares, such as $x_0,\ x_1,\ \ldots,\ x_t\in \mathbb{Z}_q$. However, for arithmetic masking the relation between $x$ and $(t+1)$ shares of $x$ is $x = (x_0 + x_1 + \cdots + x_t) \bmod q $, and in Boolean masking the relation between $x$ and $(t+1)$ shares of $x$ is $x = (x_0 \xor x_1 \xor \cdots \xor x_t)$.

\subsection{Arithmetic Operations}

It can be seen from Figure \ref{fig:saberdecaps} that the decapsulation algorithm of each KEM of the suite Scabbard consists of mostly arithmetic operations, such as polynomial multiplications, polynomial addition, and polynomial subtractions. These operations can be masked efficiently utilizing arithmetic masking. Here, we need to duplicate these operations for each arithmetic share and perform them separately. The performance cost of these operations grows linearly with the increase of arithmetic shares. 

Although this part is more or less similar for all the LWE/LWR-based KEMs (for example, Kyber and Saber), the parameter set impacts the performance of unmasked and masked versions of these operations. This also helps the schemes of Scabbard to achieve better performance compared to other LBC-based KEMs in some scenarios. The performance cost of the masked arithmetic operations in Sable is less than Saber or Kyber because the total cost of arithmetic operations of Sable is less than Saber or Kyber in the unmasked domain. It happens because Sable uses a slightly reduced parameter set than Saber. However, the performance cost of arithmetic operations in Florete or Espada is more than Saber or Kyber, as is the case in the unmasked domain.

\subsection{Compression}

Compression operation is the final step of the \texttt{LWR.PKE.Dec} algorithm, and in this step, encoded message bits are retrieved from the polynomial $m''$ after performing the reconciliation. 
For Florete and Sable, only the most significant bit is extracted, and for Espada, the four most significant bits are extracted from each coefficient of the polynomial $m''$. After that, these message bits are used as input in \texttt{SHA3-512} hash function for computing the seed $s'$ for the re-encryption procedure. These message bits are also needed to construct the session key. 
The extraction of the most significant bits is performed by using a logical shift operation in LWR-based KEM. This operation is easy to protect with Boolean masking. However, in the masked setting, the input of the compression operation is arithmetically masked, as its previous steps consisted of arithmetic operations. So, in the masked compression operation, first, we apply arithmetic to Boolean (\texttt{A2B}) conversion, and then we perform coefficient-wise $\epsilon_p - B$ bit right shift operation~\cite{HO_mask_Saber}. 

This compress operation in Sable is very similar to the one used in Saber, except for the value of $\epsilon_p$. The value of the parameter $\epsilon_p$ is smaller in Sable than in Saber. So, the performance of \texttt{A2B} conversion is relatively better in Sable compared to Saber. Hence, the overall performance of the masked compress operation is better in Sable than in Saber. The compress operation of Florete is also similar to the compress operation used in Saber. The value of parameters $\epsilon_p$ in Florete is the same as Sable and so a little smaller than in Saber. However, the degree of the message containing part of the ciphertext polynomial is $768$ in Florete, while it is $256$ in Saber. So, the number of coefficients in Florete is three times compared to Saber. The performance cost of \texttt{A2B} conversion and $\epsilon_p-1$ right shift operation in Florete is approximately three times the performance cost of these operations in Saber. Therefore, the performance of the masked compress operation in Florete takes approximately three times the cycles compared to the masked compress operation in Saber. The scheme Espada encodes four message bits in a single coefficient of ciphertext, and the polynomial size in Espada is $64$, which is $1/4$th of the polynomial size in Saber. The value of $\epsilon_p$ in Espada is slightly bigger than in Saber. However, the \texttt{A2B} conversion component is faster in Espada than in Saber due to the small polynomial size. Also, for the same reason, the coefficient-wise $\epsilon_p - 4$ bit right shift operation in Espada is faster than the coefficient-wise $\epsilon_p - 1$ bit right shift operation of Saber. Overall, the performance of the masked compress operation of Espada is roughly four times faster compared to the masked compress operation in Saber. As Kyber uses prime moduli, the masked compress operation of Kyber is far more complicated and has some extra steps. These extra steps includes conversion of arithmetic shares from $\mathbb{Z}_q$ to power-of-two modulus $\mathbb{Z}_{2^{k_q}}$, where $\log\ {q} < {2^{k_q}}$. These are computationally quite expensive operations. Due to the power-of-two moduli, schemes in Scabbard and Saber do not need these additional steps. This results in more efficient masked compress operation for these schemes.

\subsection{Message Decoding and Encoding}

For Florete and Espada, the bit length of the message \textit{i.e} 256 is not equal to the sizes of the polynomial ring, which are $768$ and $64$, respectively. Authors of Scabbard proposed techniques to encode and decode the message into the polynomial named $\mathtt{arrange\_msg}$ and $\mathtt{original\_msg}$ respectively. The encoding and decoding operation where the polynomial ring length is the same as the message length is very straightforward, and we do not need any special masking gadget for $\mathtt{original\_msg}$ and $\mathtt{arrange\_msg}$ functions. However, we need to use a special masking component to mask the $\mathtt{original\_msg}$ function when polynomial length equals $r$ times message bits, where $r>1$, e.g., Florete, NewHope~\cite{DBLP:conf/uss/AlkimDPS16}. We use $r$ coefficients to hide one message bit in this case. We also have to use a special masking gadget to mask the $\mathtt{arrange\_msg}$ function if the number of message bits equals $B$ times a polynomial length, where $B>1$, e.g., Espada. In these schemes, $B$ message bits are hidden in a coefficient. We discuss these gadgets below.

\subsubsection*{Message Decoding:} 
In Florete, $3$ coefficients had been used to hide one message bit. The $\mathtt{original\_msg}: \mathbb{Z}_2^{768} \longrightarrow \mathbb{Z}_2^{256} $ is defined here as if $\mathtt{original\_msg}(m'') = m'$ and $b\in \{ 0,\ 1,\ \ldots,\ 255 \}$ then
$m'[b] =
\begin{cases}
0 & \text{if } m''[b]+m''[b+256]+m''[b+512]\leq 1 \\
1 & \text{otherwise } \\
\end{cases}\,.
$
First, we perform secure additions (\texttt{SecAdd}) over Boolean shared data to mask this function, and the possible output must be one of $\{0,\ 1,\ 2,\ 3\}$. Notice that it is always a two-bit number for any bit $b$. The output of the $\mathtt{original\_msg}$ is equal to the most significant bit, which is the $2$nd bit. So, after performing the masked addition, we extract the most significant bit of the masked output shares ($2$nd bit). At last, we return the most significant bit as output $\mathtt{original\_msg}$ for each bit $b\in \{ 0,\ 1,\ \ldots,\ 255 \}$. We present this masked function in Algorithm \ref{algo:OMsg_florete}.

\begin{algorithm}[!ht]
\caption{ Masked $\mathtt{original\_msg}$ function for Florete }
\label{algo:OMsg_florete}
\Input{$\{m''_i\}_{1\leq i \leq n}$ where $m''_i \in \mathbb{Z}_2^{768}$ such that $\bigoplus_{i=1}^{n} m''_i = m''$}
\Output{$\{m'_i\}_{1\leq i \leq n}$ where $m'_i \in \mathbb{Z}_2^{256}$, $\bigoplus_{i=1}^{n} m'_i = m'$ and $\mathtt{original\_msg}(m'') = m'$}
\BlankLine
\For{\texttt{j=0 to 255}}
{
$\{x_i[j]\}_{1\leq i \leq n} \leftarrow m''_i[j];$  $\{y_i[j]\}_{1\leq i \leq n} \leftarrow m''_i[256+j];$  $\{z_i[j]\}_{1\leq i \leq n} \leftarrow m''_i[512+j]$\\
}

$\{w_i\}_{1\leq i \leq n} \leftarrow \mathtt{SecAdd}(\{x_i\}_{1\leq i \leq n},\{y_i\}_{1\leq i \leq n})$\\
$\{w'_i\}_{1\leq i \leq n} \leftarrow \mathtt{SecAdd}(\{w_i\}_{1\leq i \leq n},\{z_i\}_{1\leq i \leq n})$\\

$\{m'_i\}_{1\leq i \leq n} \leftarrow \{w'_i\}_{1\leq i \leq n} \gg 1$

\algorithmicreturn{ $\{m'_i\}_{1\leq i \leq n}$}
\end{algorithm}

\subsubsection*{Message Encoding:} 
In Florete and Sable, a co-efficient of the message polynomial carries a single message bit. Here, $\mathtt{arrange\_msg}$ is defined by $\mathtt{arrange\_msg}: \mathbb{Z}_2^{256} \longrightarrow \mathbb{Z}_2^{768}$ and $\mathtt{arrange\_msg}: \mathbb{Z}_2^{256} \longrightarrow \mathbb{Z}_2^{256}$ for Florete and Sable respectively. The Boolean masked output of this function then takes part in the modular addition in the next step of the re-encryption stage as the message polynomial. 
As the shares of each coefficient of the message polynomial are in $\mathbb{Z}_2$, the Boolean shares are equivalent to the arithmetic shares. Hence, we can skip the Boolean to arithmetic conversion here.
However, for Espada, we encode four message bits in a single co-efficient of the message polynomial, and $\mathtt{arrange\_msg}$ is defined by $\mathtt{arrange\_msg}: \mathbb{Z}_2^{256} \longrightarrow \mathbb{Z}_4^{64}$. So, we need to convert Boolean shares of each coefficient of message polynomial to arithmetic shares using the \texttt{B2A} algorithm. After that, we perform the modular addition with two arithmetically masked inputs.  
 
\subsection{Hash Functions}\label{hash}

Decapsulation algorithm uses one hash functions $G$ (\texttt{SHA3-512}) and one pseudo-random number generator $XOF$ (\texttt{SHAKE-128}). These functions are different instances of the sponge function Keccak-f[1600] \cite{keccak_ec}. It consists of five steps: (i) $\theta$, (ii) $\rho$, (iii) $\pi$, (iv) $\chi$, and (v) $\iota$. Among the five steps, $\theta$, $\rho$, and $\pi$ are linear diffusion steps and $\iota$ is a simple addition. As all these four steps are linear operations over Boolean shares, in masked settings, we repeat all these operations on each share separately. Only $\chi$ is a degree 2 non-linear mapping and thus requires extra attention to mask. Overall, Keccak-f[1600] is less expensive to mask by using Boolean masking. %Similar to higher-order masked Saber \cite{HO_mask_Saber}, 
Here, we use the higher-order masked Keccak proposed by Gross et al.~\cite{ho-masked-keccak}. 
Due to the compact parameter choices, Scabbard schemes require fewer pseudo-random numbers than Saber. Eventually, this leads to fewer invocations of the sponge function Keccak in Florete and Sable than in Espada. Moreover, the output length of \texttt{SHAKE-128} is the same for Florete and Sable, which is even smaller than Espada. To sum up, the performance cost of the masked XOF \texttt{SHAKE-128} is lower in Florete, Sable, and Espada compared to Saber.

\subsection{Centered Binomial Sampler}\label{cbd_sampler}

The re-encryption part of the decapsulation algorithm contains a centered binomial sampler for sampling the vector $\mathbf{s'}$. This sampler outputs $\mathtt{HW}(x) - \mathtt{HW}(y)$, where $x$ and $y$ are pseudo-random numbers and $\mathtt{HW}$ represents hamming weight. The bit size of pseudo-random numbers $x$ and $y$ depends on the scheme. These pseudo-random numbers are produced employing \texttt{SHAKE-128}. As mentioned in the previous section, these function is efficient if we mask with the help of Boolean masking. Hence, the shares generated from \texttt{SHAKE-128} are Boolean. However, upon constructing the $\mathbf{s'}$, we need to perform modular multiplication with inputs $\mathbf{s'}$ and public-key $\mathbf{b}$. This is efficient if we use arithmetic masking. Therefore, we need to perform Boolean to arithmetic conversion in the masked-centered binomial sampler. %for converting Boolean shares to arithmetic shares.  
Schneider et al.~\cite{Tobias2019_binomial_sampler} proposed two centered binomial samplers, $\mathtt{Sampler}_1$ and $\mathtt{Sampler}_2$. $\mathtt{Sampler}_1$ first converts Boolean shares of $x$ and $y$ to arithmetic shares then computes $\mathtt{HW}(x) - \mathtt{HW}(y)$ by using arithmetic masking technique. $\mathtt{Sampler}_2$ first computes $z = \mathtt{HW}(x) - \mathtt{HW}(y) + k$, where $k \geq \mu/2$ using Boolean masking. After that, it converts Boolean shares of $z$  to arithmetic shares and then performs $z-k$ using the arithmetic masking technique to remain with arithmetic shares of $\mathtt{HW}(x) - \mathtt{HW}(y)$. $\mathtt{Sampler}_1$ uses a bit-wise masking procedure, while $\mathtt{sampler}_2$ uses the bitslicing technique on some parts of the algorithm for receiving better throughput. 
We have adopted these two samplers and optimized them to mask the CBD function of each KEM of the Scabbard suite.
We could not directly use the optimized CBD used in Saber~\cite{HO_mask_Saber}, as that one is optimized for $\beta_8$, and schemes of Scabbard use smaller CBD to sample the vector $\mathbf{s'}$. Schemes like Kyber and NewHope \cite{DBLP:conf/uss/AlkimDPS16,Tobias2019_binomial_sampler} use prime modulus. So, a few components there are different, for example, the B2A conversion and extra modular addition. As Scabbard uses power-of-two moduli, these components can be implemented in a much cheaper way for them. 
We describe the optimized masked CBD samplers for these schemes below.

\subsubsection{Florete and Sable:}

In these two schemes, we take advantage of the centered binomial sampler with a small standard deviation, $\beta_2$. For $\beta_2$, $x$ and $y$ are $1$-bit pseudo-random numbers. We have adopted $\mathtt{Sampler}_1$ and $\mathtt{Sampler}_2$, with these specification. As $\mathtt{Sampler}_2$ is designed to provide a better performance, we started with the adaptation of $\mathtt{Sampler}_2$ for $\beta_2$ named $\mathtt{MaskCBDSampler}_A$ as shown in Algorithm~\ref{algo:SecSampler2_Florete_Sable}. In this algorithm, first, we perform \texttt{SecBitSub} on Boolean shares of $x$ and $y$ to calculate Boolean shares of $\mathtt{HW}(x) - \mathtt{HW}(y)$. Second, we add constant $1$ with the output shares of \texttt{SecBitSub} to avoid negative numbers. Third, we convert the output from Boolean shares to arithmetic shares with the help of the \texttt{B2A} conversion algorithm proposed in~\cite{Bettale2018_B2A}. In the last step, we subtract the added constant in step-$2$, which converts secret shares from $\{0,1,2\}$ to $\{-1,0,1\}$. 

\begin{algorithm}[!ht]
\caption{ $\mathtt{MaskCBDSampler}_A$ (\cite{Tobias2019_binomial_sampler}, using $\mathtt{sampler}_2$) }
\label{algo:SecSampler2_Florete_Sable}
\Input{$\{x_i\}_{0\leq i \leq n},\{y_i\}_{0\leq i \leq n}$ where $x_i,y_i \in \mathbb{R}_2$ such that $\bigoplus_{i=0}^{n} x_i = x, \bigoplus_{i=0}^{n} y_i = y$}
\Output{$\{A_i\}_{0\leq i \leq n}$ where  $A_i \in \mathbb{R}_q$ and $\sum_{i=0}^{n} A_i = (\mathtt{HW}(x)- \mathtt{HW}(y)) \bmod q $}
\BlankLine
$\{z_i\}_{0\leq i \leq n}\leftarrow$ \texttt{SecBitSub}$(\{x_i\}_{0\leq i \leq n},\{y_i\}_{0\leq i \leq n})$\\
$z_0[0] \leftarrow z_0[0] \xor 1$\\
% $\{z_i\}_{0\leq i \leq n}\leftarrow$ \texttt{SecConstAdd}$(\{z_i\}_{0\leq i \leq n}, 1)$~\ref{algo:SecConstAdd_Florete_Sable}\\
$\{A_i\}_{0\leq i \leq n}\leftarrow$ \texttt{B2A}$(\{z_i\}_{0\leq i \leq n})$~\cite{Bettale2018_B2A}\\
$A_1\leftarrow (A_1- 1) \bmod q$\\
\algorithmicreturn{ $\{A_i\}_{0\leq i \leq n}$}
\end{algorithm}

\begin{algorithm}[!ht]
\caption{ $\mathtt{MaskCBDSampler}_B$ (\cite{Tobias2019_binomial_sampler}, using $\mathtt{sampler}_1$) }
\label{algo:SecSampler1_Florete_Sable}

\Input{$\{x_i\}_{0\leq i \leq n},\{y_i\}_{0\leq i \leq n}$ where $x_i,y_i \in \mathbb{R}_2$ such that $\bigoplus_{i=0}^{n} x_i = x, \bigoplus_{i=0}^{n} y_i = y$}
\Output{$\{A_i\}_{0\leq i \leq n}$ where  $A_i \in \mathbb{R}_q$ and $\sum_{i=0}^{n} A_i = (\mathtt{HW}(x)- \mathtt{HW}(y)) \bmod q $}
\BlankLine
$\{T1_i\}_{0\leq i \leq n}\leftarrow$ \texttt{B2A}$(\{x_i\}_{0\leq i \leq n})$~\cite{Bettale2018_B2A};\ \ $\{T2_i\}_{0\leq i \leq n}\leftarrow$ \texttt{B2A}$(\{y_i\}_{0\leq i \leq n})$~\cite{Bettale2018_B2A}\\
\For{\texttt{i=0 to n}}
{
$A_i\leftarrow (T1_i - T2_i)$\\
}
\algorithmicreturn{ $\{A_i\}_{0\leq i \leq n}$}
\end{algorithm}

As the bit size of $x$ and $y$ is small for $\beta_2$, the bitslice technique for addition and subtraction does not improve the throughput much. So, for comparison purposes, we have adopted the technique of the $\mathtt{sampler}_1$ for $\beta_2$. We name this algorithm $\mathtt{MaskCBDSampler}_A$, and present in Algorithm~\ref{algo:SecSampler1_Florete_Sable}. In this algorithm, we conduct \texttt{B2A} conversions over $x$ and $y$ and then perform share-wise subtraction between arithmetic shares of $x$ and $y$. 

\begin{algorithm}[t]
\caption{ $\mathtt{MaskCBDSampler}_C$ (\cite{Tobias2019_binomial_sampler}, using $\mathtt{sampler}_2$) }
\label{algo:SecSampler2_Espada}

\Input{$\{x_i\}_{0\leq i \leq n},\{y_i\}_{0\leq i \leq n}$ where $x_i,y_i \in \mathbb{R}_2^3$ such that $\bigoplus_{i=0}^{n} x_i = x, \bigoplus_{i=0}^{n} y_i = y$}
\Output{$\{A_i\}_{0\leq i \leq n}$ where  $A_i \in \mathbb{R}_q$ and $\sum_{i=0}^{n} A_i = (\mathtt{HW}(x)- \mathtt{HW}(y)) \bmod q $}
\BlankLine
$\{z_i\}_{0\leq i \leq n}\leftarrow$ \texttt{SecBitAdd}$(\{x_i\}_{0\leq i \leq n})$~\cite{FO-masked-saber} \\
$\{z_i\}_{0\leq i \leq n}\leftarrow$ \texttt{SecBitSub}$(\{z_i\}_{0\leq i \leq n},\{y_i\}_{0\leq i \leq n})$~\cite{Tobias2019_binomial_sampler} \\
\For{\texttt{i=0 to n}}
{
$z_i[2]\leftarrow (z_i[2] \xor z_i[1])$\\
}
$z_0[2]\leftarrow z_0[2]\xor 1$\\
% $\{z_i\}_{0\leq i \leq n}\leftarrow$ \texttt{SecConstAdd}$(\{z_i\}_{0\leq i \leq n},4)$~\ref{algo:SecConstAdd_Espada}\\
$\{A_i\}_{0\leq i \leq n}\leftarrow$ \texttt{B2A}$(\{z_i\}_{0\leq i \leq n})$~\cite{Bettale2018_B2A}\\
$A_1\leftarrow (A_1- 4) \bmod q$\\

\algorithmicreturn{ $\{A_i\}_{0\leq i \leq n}$}
\end{algorithm}

\subsubsection{Espada:}

We use the centered binomial sampler, $\beta_6$, in this scheme. For $\beta_6$, $x$ and $y$ are $3$-bit pseudo-random numbers. We have adopted a bitsliced implementation of $\mathtt{Sampler}_2$ from \cite{Tobias2019_binomial_sampler} for $\beta_6$ to achieve better efficiency as the standard deviation of the CBD is large. We name this masked sampler as $\mathtt{MaskCBDSampler}_C$, and it is shown in Algorithm~\ref{algo:SecSampler2_Espada}. Similar to $\mathtt{MaskCBDSampler}_B$, $\mathtt{MaskCBDSampler}_C$ begins with the \texttt{SecBitAdd} operation, which is performed on Boolean shares of $x$ and generates Boolean shares of $\mathtt{HW}(x)$. Then \texttt{SecBitSub} is conducted over the Boolean output shares and Boolean shares of $y$ and outputs Boolean shares of $\mathtt{HW}(x) - \mathtt{HW}(y)$. After that, the constant $4$ is added with the output shares of \texttt{SecBitSub} to avoid negative numbers. In the next step, we convert the output from Boolean shares to arithmetic shares with the help of \texttt{B2A} conversion algorithm proposed in~\cite{Bettale2018_B2A}. Finally, we subtract the added constant in step-$7$ and transform secret shares from $\{1,2,3,4,5,6,7\}$ to $\{-3,-2,-1,0,1,2,3\}$.

The masked CBD sampler ($\beta_8$) used in Saber is faster than the masked CBD of Kyber because of the power-of-two moduli. $\mathtt{MaskCBDSampler}_A$ and $\mathtt{MaskCBDSampler}_B$ are optimized implementation of $\beta_2$, which has been used in Florete and Sable. $\mathtt{MaskCBDSampler}_C$ is designed for Espada, which is optimized implementation of $\beta_6$. For $\beta_2$ and $\beta_6$, the B2A conversion is much faster than $\beta_8$ thanks to the smaller coefficients size in the input polynomial. Therefore, the performance cost of the masked CBD is less for all the schemes in Scabbard compared to Saber or Kyber. A more detailed performance cost analysis of masked CBD implementations for Scabbard is presented in Section \ref{sec:cbd_performance}.        

\subsection{Ciphertext Comparison}

It is one of the costliest components for masked implementations of lattice-based KEMs, which is a part of the FO transformation. Previously, many methods have been proposed to perform this component efficiently \cite{HO_mask_Saber,D'AnversVV22,ho-mask-comparator-kyber-bos-2021}. For the masked ciphertext comparison part of each KEM of Scabbard, we have adopted the {improved simple masked comparison} method used in the higher-order masked implementation of Saber~\cite{HO_mask_Saber}. To the best of our knowledge, this is currently the most efficient masked ciphertext comparison implementation available.  
Through this process, we compare the arithmetically masked output of the re-encryption component before the right shift operation $(\tilde{\emph{u}},\ \tilde{v})$ with the unmasked public ciphertext, ($\emph{u},\ v$). Additionally, note that $\mathbf{u}'=\tilde{\mathbf{u}}\gg (\epsilon_q-\epsilon_p)$ and $v'=\tilde{v} \gg (\epsilon_p-\epsilon_t-B)$. At first, we perform \texttt{A2B} conversion step over the arithmetically masked shares of the output and transform these to Boolean shares, and then we follow the right shift operation. After that, we subtract the unmasked public ciphertext ($\emph{u},\ v$) from a share of the Boolean masked output of the \texttt{A2B} operation with the help of the XOR operation. Finally, we proceed with checking that all the returned bits of the subtract operation are zero with the \texttt{BooleanAllBitsOneTest} algorithm. This algorithm returns $1$ only if it receives all the bits encoded in each coefficient of the polynomials is $1$; else it returns $0$. All these aforementioned steps are presented in Algorithm~\ref{algo:MaskComp1}. For further details, we refer to the higher-order masked Saber paper~\cite{HO_mask_Saber}.  

\begin{algorithm}[!ht]
\caption{ Simple masked comparison algorithm~\cite{HO_mask_Saber} }
\label{algo:MaskComp1}

\Input{Arithmetic masked re-encrypted ciphertext ($\{\tilde{\mathbf{u}}_i\}_{0\leq i \leq n}$, $\{\tilde{v}_i\}_{0\leq i \leq n}$)\\ and public ciphertext ($\mathbf{u}$ and $v$) where each $\tilde{\mathbf{u}}_i \in \mathbb{R}_{2^{\epsilon_q}}^l$ and ${\tilde{v}}_i \in \mathbb{R}_{2^{\epsilon_p}}$ \\and $\sum_{i=0}^{n} \tilde{\mathbf{u}}_i \bmod{q} = \tilde{\mathbf{u}}$ $\sum_{i=0}^{n} \tilde{v}_i \bmod{q} = \tilde{v}$.}
\Output{$\{bit\}_{0\leq i \leq n}$, where with each $bit_i \in \mathbb{Z}_2$ and $\bigoplus_{i=0}^{n} bit_i = 1$ iff\\ $\mathbf{u}=\mathbf{u}'\gg (\epsilon_q-\epsilon_p)$ and $v=v'\gg (\epsilon_p-\epsilon_t-B)$, otherwise $0$.}
\BlankLine
\BlankLine

$\{\mathbf{y}_i\}_{0\leq i \leq n}\leftarrow$ \texttt{A2B}$(\{\tilde{\mathbf{u}}_i\}_{0\leq i \leq n});$  $\{x_i\}_{0\leq i \leq n}\leftarrow$ \texttt{A2B}$(\{\tilde{v}_i\}_{0\leq i \leq n})$\\
$\{\mathbf{y}_i\}_{0\leq i \leq n}\leftarrow (\{\mathbf{y}_i\}_{0\leq i \leq n}\gg (\epsilon_q-\epsilon_p));$  $\{x_i\}_{0\leq i \leq n}\leftarrow(\{x_i\}_{0\leq i \leq n}\gg (\epsilon_p-\epsilon_t-B))$\\
$\mathbf{y}_1\leftarrow \mathbf{y}_1 \xor \mathbf{u}; $  ${x}_1\leftarrow {x}_1 \xor {v}$\\
\tcc{Boolean circuit to test all bits of $(\mathbf{y},x)$ are $0$} 
$\mathbf{y}_0\leftarrow \lnot \mathbf{y}_0; $  ${x}_0\leftarrow \lnot {x}_0$\\
$\{bit_i\}_{0\leq i \leq n}\leftarrow$ \texttt{BooleanAllBitsOneTest} $(\{\mathbf{y}_i\}_{0\leq i \leq n},\{{x}_i\}_{0\leq i \leq n},\epsilon_p,\epsilon_t)$\\
\BlankLine
\algorithmicreturn{ $\{bit_i\}_{0\leq i \leq n}$}
\end{algorithm}

\begin{table}[!ht]
\centering
\caption{Size of inputs of the \texttt{A2B} and \texttt{BooleanAllBitsOneTest} functions situated in Algorithm~\ref{algo:MaskComp1} for Scabbard's schemes and Saber}
\label{tab:function_byte}
\begin{tabular}{ccccccccc}
\hline
Function                       & \multicolumn{8}{c}{Input Bytes}                                                                                                                                                                                                      \\ \cline{2-9} 
                               & { } & { Florete} & { } & { Sable} & { } & { Espada} & { } & { Saber} \\ \hline
\texttt{A2B}                   &                         & 1824                           &                         & 1344                         &                         & 1544                          &                         & 1568                         \\ \hline
\texttt{BooleanAllBitsOneTest} &                         & 1248                           &                         & 1024                         &                         & 1304                          &                         & 1088                         \\ \hline
\end{tabular}
\end{table}

The parameter settings are different for each KEM of the Scabbard suite. Due to this, byte sizes of the masked inputs of the functions \texttt{A2B} and \texttt{BooleanAllBitsOneTest} are different for each KEM of the suite, and we show these numbers in Table~\ref{tab:function_byte}. For reference, we also provide the byte sizes of the masked inputs of \texttt{A2B} and \texttt{BooleanAllBitsOneTest} for Saber in this table. These differences in the input bytes also affect the performances of corresponding masked implementations. The masked input sizes of both the functions \texttt{A2B} and \texttt{BooleanAllBitsOneTest} for Sable are less than Saber. On account of this, the performance cost of masked ciphertext comparison is cheaper for Sable than Saber. The masked input sizes of both functions \texttt{A2B} and \texttt{BooleanAllBitsOneTest} for Florete are greater than Saber. So, the masked ciphertext comparison component of Florete needs more cycles than Saber. The masked input size of the function \texttt{A2B} of Espada is less than Saber, but the input size of \texttt{BooleanAllBitsOneTest} for Espada is bigger than Saber. So, the first-order masked comparison component is faster for Espada compared to Saber, but the second and third-order masked comparison component is slower in Espada than in Saber. However, the performance of each scheme's masked ciphertext comparison component in the suite Scabbard is better than Kyber because of the prepossessing steps needed in Kyber \cite{D'AnversVV22}. 

\section{Performance Evaluation}

We implemented all our algorithms on a 32-bit ARM Cortex-M4 microcontroller, STM32F407-DISCOVERY development board. We used the popular post-quantum cryptographic library and benchmarking framework PQM4~\cite{PQM4} for all measurements.
The system we used to measure the performance of the masked implementations includes the compiler \texttt{arm-none-eabi-gcc} version \texttt{9.2.1}. The PQM4 library uses the system clock to measure the clock cycle, and the frequency of this clock is 24MHz. 
We employ random numbers to ensure the independence of the shares of the masked variable in masking algorithms. For this purpose, we use the on-chip \texttt{TRNG} (true random number generator) of the ARM Cortex-M4 device. This \texttt{TRNG} has a different clock frequency than the main system clock, which is 48MHz. It generates a $32$-bit random number in $40$ clock cycles, equivalent to $20$ clock cycles for the main system clock. Our implementations can be used for any order of masking. In this section, we provide the performance details of first-, second-, and third-order masking. 

%-----------------original table--------------

\begin{table}[!ht]
\centering
\caption{Performance of $\mathtt{MaskCBDSampler}_A$ and $\mathtt{MaskCBDSampler}_B$}
\label{tab:table-sampler}

\begin{tabular}{lccccccc}
\hline
\multicolumn{1}{c}{{ }}                          & \multicolumn{7}{c}{{ x1000 clock cycles}}                            \\ \cline{2-8} 
\multicolumn{1}{c}{{{ Order}}} &  & 1st              &           & 2nd              &  & 3rd              &  \\ \hline
$\mathtt{MaskCBDSampler}_A$                                          &  & \textbf{178,591} & \textbf{} & 504,101          &  & 1,226,224        &  \\
$\mathtt{MaskCBDSampler}_B$                                          &  & 182,714          &           & \textbf{499,732} &  & \textbf{909,452} &  \\ \hline
\end{tabular}
\end{table}

%----------------------original table ends--------------------

\subsection{Analyzing the Performance of Masked CBD Samplers}\label{sec:cbd_performance}

As discussed in Section \ref{cbd_sampler}, $\mathtt{MaskCBDSampler}_A$ and $\mathtt{MaskCBDSampler}_B$ can be used for both Florete and Sable. Performance comparisons between $\mathtt{MaskCBDSampler}_A$ and $\mathtt{MaskCBDSampler}_B$ for different shares are provided in Table \ref{tab:table-sampler}. Overall, we observe from the table that $\mathtt{MaskCBDSampler}_B$ performs better than $\mathtt{MaskCBDSampler}_A$ for higher-order masking. As a result, we use $\mathtt{MaskCBDSampler}_B$ in the masked implementations of Florete and Sable.

\subsection{Performance Measurement of Masked Scabbard Suite}

Table \ref{tab:Florete_performance}, \ref{tab:Espada_performance}, and \ref{tab:Sable_performance} provide the clock cycles required to execute the masked decapsulation algorithm of Florete, Espada, and Sable, respectively. The overhead factors for the first-, second-, and third-order masked decapsulation operation of Florete are $2.74$x, $5.07$x, and $7.75$x compared to the unmasked version. For Espada, the overhead factors for the first-, second-, and third-order decapsulation algorithm compared to the unmasked decapsulation are $1.78$x, $2.82$x, and $4.07$, respectively. Similarly, for Sable, the overhead factors for the first-, second-, and third-order decapsulation algorithm are $2.38$x, $4.26$x, and $6.35$x than the unmasked one. As mentioned earlier, the masked algorithm needs fresh random numbers to maintain security. Generating random numbers is a costly procedure. So, for a better understanding of the improvements, we also present the requirement of random bytes for Florete, Espada, and Sable in Table~\ref{tab:Scabbard_random}.

\begin{table}[!ht]
\centering
% \scriptsize
\caption{Performance of Florete}
\label{tab:Florete_performance}
\resizebox{.9\textwidth}{!}{%
\begin{tabular}{llllllllrrrrrrrrrr}
\hline
                      &  & \multicolumn{5}{l}{}                                                  &  & \multicolumn{10}{c}{x1000 clock cycles}                                                                                                                                                                                                                                                                                            \\ \cline{9-18} 
                      &  & \multicolumn{5}{c}{Order}                                             &  & \multicolumn{1}{c|}{Unmask}                & \multicolumn{3}{c|}{1st}                                                                       & \multicolumn{3}{c|}{2nd}                                                                         & \multicolumn{3}{c}{3rd}                                                      \\ \hline
\multicolumn{7}{l}{{ \textbf{Florete CCA-KEM-Decapsulation}}}                &  & \multicolumn{1}{r|}{954}                   &                      & 2,621                 & \multicolumn{1}{r|}{(2.74x)}                    &                      & 4,844                   & \multicolumn{1}{r|}{(5.07x)}                    &                      & 7,395                   & (7.75x)                     \\
\multicolumn{1}{l|}{} &  & \multicolumn{5}{l}{\textbf{CPA-PKE-Decryption}}                       &  & \multicolumn{1}{r|}{248}                   &                      & 615                   & \multicolumn{1}{r|}{(2.47x)}                    &                      & 1,107                   & \multicolumn{1}{r|}{(4.46x)}                    &                      & 1,651                   & (6.65x)                     \\
\multicolumn{1}{l|}{} &  & \multicolumn{1}{l|}{} &  & \multicolumn{3}{l}{Polynomial arithmetic}  &  & \multicolumn{1}{r|}{241}                   &                      & 461                   & \multicolumn{1}{r|}{(1.91x)}                    &                      & 690                     & \multicolumn{1}{r|}{(2.86x)}                    &                      & 917                     & (3.80x)                     \\
\multicolumn{1}{l|}{} &  & \multicolumn{1}{l|}{} &  & \multicolumn{3}{l}{Compression}            &  & \multicolumn{1}{r|}{}                      &                      &                       & \multicolumn{1}{r|}{}                           &                      &                         & \multicolumn{1}{r|}{}                           &                      &                         &                             \\
\multicolumn{1}{l|}{} &  & \multicolumn{1}{l|}{} &  & \multicolumn{3}{l}{$original\_msg$}        &  & \multicolumn{1}{r|}{\multirow{-2}{*}{6}}   & \multicolumn{1}{l}{} & \multirow{-2}{*}{153} & \multicolumn{1}{r|}{\multirow{-2}{*}{(25.50x)}} & \multicolumn{1}{l}{} & \multirow{-2}{*}{416}   & \multicolumn{1}{r|}{\multirow{-2}{*}{(69.33x)}} & \multicolumn{1}{l}{} & \multirow{-2}{*}{734}   & \multirow{-2}{*}{(122.33x)} \\
\multicolumn{1}{l|}{} &  & \multicolumn{5}{l}{\textbf{Hash $\mathcal{G}$ (SHA3-512)}}            &  & \multicolumn{1}{r|}{13}                    &                      & 123                   & \multicolumn{1}{r|}{(9.46x)}                    &                      & 242                     & \multicolumn{1}{r|}{(18.61x)}                   &                      & 379                     & (29.15x)                    \\
\multicolumn{1}{l|}{} &  & \multicolumn{5}{l}{\textbf{CPA-PKE-Encryption}}                       &  & \multicolumn{1}{r|}{554}                   &                      & 1,744                 & \multicolumn{1}{r|}{(3.14x)}                    &                      & 3,354                   & \multicolumn{1}{r|}{(6.05x)}                    &                      & 5,225                   & (9.43x)                     \\
\multicolumn{1}{l|}{} &  & \multicolumn{1}{l|}{} &  & \multicolumn{3}{l}{Secret generation}      &  & \multicolumn{1}{r|}{29}                    &                      & 427                   & \multicolumn{1}{r|}{(14.72x)}                   &                      & 982                     & \multicolumn{1}{r|}{(33.86x)}                   &                      & 1,663                   & (57.34x)                    \\
\multicolumn{1}{l|}{} &  & \multicolumn{1}{l|}{} &  & \multicolumn{1}{l|}{} &  & XOF (SHAKE-128) &  & \multicolumn{1}{r|}{25}                    &                      & 245                   & \multicolumn{1}{r|}{(9.80x)}                    &                      & 484                     & \multicolumn{1}{r|}{(19.36x)}                   &                      & 756                     & (30.24x)                    \\
\multicolumn{1}{l|}{} &  & \multicolumn{1}{l|}{} &  & \multicolumn{1}{l|}{} &  & CBD ($\beta_2$) &  & \multicolumn{1}{r|}{4}                     &                      & 182                   & \multicolumn{1}{r|}{(45.50x)}                   &                      & 497                     & \multicolumn{1}{r|}{(124.25x)}                  &                      & 907                     & (226.75x)                   \\
\multicolumn{1}{l|}{} &  & \multicolumn{1}{l|}{} &  & \multicolumn{3}{l}{Polynomial arithmetic}  &  & \multicolumn{1}{r|}{}                      &                      &                       & \multicolumn{1}{r|}{}                           &                      &                         & \multicolumn{1}{r|}{}                           &                      &                         &                             \\
\multicolumn{1}{l|}{} &  & \multicolumn{1}{l|}{} &  & \multicolumn{3}{l}{$arrange\_msg$}         &  & \multicolumn{1}{r|}{}                      & \multicolumn{1}{l}{} & \multirow{-2}{*}{943} & \multicolumn{1}{r|}{}                           & \multicolumn{1}{l}{} & \multirow{-2}{*}{1,357} & \multicolumn{1}{r|}{}                           & \multicolumn{1}{l}{} & \multirow{-2}{*}{1,783} &                             \\
\multicolumn{1}{l|}{} &  & \multicolumn{1}{l|}{} &  & \multicolumn{3}{l}{Polynomial Comparison}  &  & \multicolumn{1}{r|}{\multirow{-3}{*}{524}} & \multicolumn{1}{c}{} & 373                   & \multicolumn{1}{r|}{\multirow{-3}{*}{(2.51x)}}  & \multicolumn{1}{l}{} & 1,014                   & \multicolumn{1}{r|}{\multirow{-3}{*}{(4.52x)}}  & \multicolumn{1}{l}{} & 1,778                   & \multirow{-3}{*}{(6.79x)}   \\
\multicolumn{1}{l|}{} &  & \multicolumn{5}{l}{Other operations}                                  &  & \multicolumn{1}{r|}{138}                   &                      & 139                   & \multicolumn{1}{r|}{(1.00x)}                    &                      & 140                     & \multicolumn{1}{r|}{(1.01x)}                    &                      & 140                     & (1.01x)                     \\ \hline
\end{tabular}%
}

\end{table}

\begin{table}[!ht]
\centering
\caption{Performance of Espada}
\label{tab:Espada_performance}
\resizebox{.9\textwidth}{!}{%
\begin{tabular}{llllllllrrrrrrrrrr}
\hline
                      &  & \multicolumn{5}{l}{}                                                  &  & \multicolumn{10}{c}{x1000 clock cycles}                                                                                                                                                                                                                                                                                                 \\ \cline{9-18} 
                      &  & \multicolumn{5}{c}{Order}                                             &  & \multicolumn{1}{c|}{Unmask}                  & \multicolumn{3}{c|}{1st}                                                                         & \multicolumn{3}{c|}{2nd}                                                                          & \multicolumn{3}{c}{3rd}                                                      \\ \hline
\multicolumn{7}{l}{{ \textbf{Espada CCA-KEM-Decapsulation}}}                 &  & \multicolumn{1}{r|}{2,422}                   &                      & 4,335                   & \multicolumn{1}{r|}{(1.78x)}                    &                      & 6,838                   & \multicolumn{1}{r|}{(2.82x)}                     &                      & 9,861                   & (4.07x)                     \\
\multicolumn{1}{l|}{} &  & \multicolumn{5}{l}{\textbf{CPA-PKE-Decryption}}                       &  & \multicolumn{1}{r|}{70}                      &                      & 137                     & \multicolumn{1}{r|}{(1.95x)}                    &                      & 230                     & \multicolumn{1}{r|}{(3.28x)}                     &                      & 324                     & (4.62x)                     \\
\multicolumn{1}{l|}{} &  & \multicolumn{1}{l|}{} &  & \multicolumn{3}{l}{Polynomial arithmetic}  &  & \multicolumn{1}{r|}{69}                      &                      & 116                     & \multicolumn{1}{r|}{(1.68x)}                    &                      & 170                     & \multicolumn{1}{r|}{(2.46x)}                     &                      & 225                     & (3.26x)                     \\
\multicolumn{1}{l|}{} &  & \multicolumn{1}{l|}{} &  & \multicolumn{3}{l}{Compression}            &  & \multicolumn{1}{r|}{}                        &                      &                         & \multicolumn{1}{r|}{}                           &                      &                         & \multicolumn{1}{r|}{}                            &                      &                         &                             \\
\multicolumn{1}{l|}{} &  & \multicolumn{1}{l|}{} &  & \multicolumn{3}{l}{$original\_msg$}        &  & \multicolumn{1}{r|}{\multirow{-2}{*}{0.4}}   & \multicolumn{1}{l}{} & \multirow{-2}{*}{20}    & \multicolumn{1}{r|}{\multirow{-2}{*}{(50.00x)}} & \multicolumn{1}{l}{} & \multirow{-2}{*}{60}    & \multicolumn{1}{r|}{\multirow{-2}{*}{(150.00x)}} & \multicolumn{1}{l}{} & \multirow{-2}{*}{99}    & \multirow{-2}{*}{(247.50x)} \\
\multicolumn{1}{l|}{} &  & \multicolumn{5}{l}{\textbf{Hash $\mathcal{G}$ (SHA3-512)}}            &  & \multicolumn{1}{r|}{13}                      &                      & 123                     & \multicolumn{1}{r|}{(9.46x)}                    &                      & 243                     & \multicolumn{1}{r|}{(18.69x)}                    &                      & 379                     & (29.15x)                    \\
\multicolumn{1}{l|}{} &  & \multicolumn{5}{l}{\textbf{CPA-PKE-Encryption}}                       &  & \multicolumn{1}{r|}{2,215}                   &                      & 3,950                   & \multicolumn{1}{r|}{(1.78x)}                    &                      & 6,240                   & \multicolumn{1}{r|}{(2.81x)}                     &                      & 9,031                   & (4.07x)                     \\
\multicolumn{1}{l|}{} &  & \multicolumn{1}{l|}{} &  & \multicolumn{3}{l}{Secret generation}      &  & \multicolumn{1}{r|}{57}                      &                      & 748                     & \multicolumn{1}{r|}{(13.12x)}                   &                      & 1,650                   & \multicolumn{1}{r|}{(28.94x)}                    &                      & 3,009                   & (52.78x)                    \\
\multicolumn{1}{l|}{} &  & \multicolumn{1}{l|}{} &  & \multicolumn{1}{l|}{} &  & XOF (SHAKE-128) &  & \multicolumn{1}{r|}{51}                      &                      & 489                     & \multicolumn{1}{r|}{(9.58x)}                    &                      & 968                     & \multicolumn{1}{r|}{(18.98x)}                    &                      & 1,510                   & (29.60x)                    \\
\multicolumn{1}{l|}{} &  & \multicolumn{1}{l|}{} &  & \multicolumn{1}{l|}{} &  & CBD ($\beta_6$) &  & \multicolumn{1}{r|}{6}                       &                      & 259                     & \multicolumn{1}{r|}{(43.16x)}                   &                      & 681                     & \multicolumn{1}{r|}{(113.50x)}                   &                      & 1,498                   & (249.66x)                   \\
\multicolumn{1}{l|}{} &  & \multicolumn{1}{l|}{} &  & \multicolumn{3}{l}{Polynomial arithmetic}  &  & \multicolumn{1}{r|}{}                        &                      &                         & \multicolumn{1}{r|}{}                           &                      &                         & \multicolumn{1}{r|}{}                            &                      &                         &                             \\
\multicolumn{1}{l|}{} &  & \multicolumn{1}{l|}{} &  & \multicolumn{3}{l}{$arrange\_msg$}         &  & \multicolumn{1}{r|}{}                        & \multicolumn{1}{l}{} & \multirow{-2}{*}{2,865} & \multicolumn{1}{r|}{}                           & \multicolumn{1}{l}{} & \multirow{-2}{*}{3,593} & \multicolumn{1}{r|}{}                            & \multicolumn{1}{l}{} & \multirow{-2}{*}{4,354} &                             \\
\multicolumn{1}{l|}{} &  & \multicolumn{1}{l|}{} &  & \multicolumn{3}{l}{Polynomial Comparison}  &  & \multicolumn{1}{r|}{\multirow{-3}{*}{2,157}} & \multicolumn{1}{c}{} & 259                     & \multicolumn{1}{r|}{\multirow{-3}{*}{(1.44x)}}  & \multicolumn{1}{l}{} & 996                     & \multicolumn{1}{r|}{\multirow{-3}{*}{(2.12x)}}   & \multicolumn{1}{l}{} & 1,667                   & \multirow{-3}{*}{(2.79x)}   \\
\multicolumn{1}{l|}{} &  & \multicolumn{5}{l}{Other operations}                                  &  & \multicolumn{1}{r|}{124}                     &                      & 124                     & \multicolumn{1}{r|}{(1.00x)}                    &                      & 124                     & \multicolumn{1}{r|}{(1.00x)}                     &                      & 126                     & (1.01x)                     \\ \hline
\end{tabular}%
}
\end{table}

\begin{table}[!ht]
\centering
\caption{Performance of Sable}
\label{tab:Sable_performance}
\resizebox{.9\textwidth}{!}{%
\begin{tabular}{llllllllrrrrrrrrrr}
\hline
                      &  & \multicolumn{5}{l}{}                                                  &  & \multicolumn{10}{c}{x1000 clock cycles}                                                                                                                                                                                                                                                                                              \\ \cline{9-18} 
                      &  & \multicolumn{5}{c}{Order}                                             &  & \multicolumn{1}{c|}{Unmask}                & \multicolumn{3}{c|}{1st}                                                                         & \multicolumn{3}{c|}{2nd}                                                                         & \multicolumn{3}{c}{3rd}                                                      \\ \hline
\multicolumn{7}{l}{{ \textbf{Sable CCA-KEM-Decapsulation}}}                  &  & \multicolumn{1}{r|}{1,020}                 &                      & 2,431                   & \multicolumn{1}{r|}{(2.38x)}                    &                      & 4,348                   & \multicolumn{1}{r|}{(4.26x)}                    &                      & 6,480                   & (6.35x)                     \\
\multicolumn{1}{l|}{} &  & \multicolumn{5}{l}{\textbf{CPA-PKE-Decryption}}                       &  & \multicolumn{1}{r|}{130}                   &                      & 291                     & \multicolumn{1}{r|}{(2.23x)}                    &                      & 510                     & \multicolumn{1}{r|}{(3.92x)}                    &                      & 745                     & (5.73x)                     \\
\multicolumn{1}{l|}{} &  & \multicolumn{1}{l|}{} &  & \multicolumn{3}{l}{Polynomial arithmetic}  &  & \multicolumn{1}{r|}{128}                   &                      & 238                     & \multicolumn{1}{r|}{(1.85x)}                    &                      & 350                     & \multicolumn{1}{r|}{(2.73x)}                    &                      & 465                     & (3.63x)                     \\
\multicolumn{1}{l|}{} &  & \multicolumn{1}{l|}{} &  & \multicolumn{3}{l}{Compression}            &  & \multicolumn{1}{r|}{}                      &                      &                         & \multicolumn{1}{r|}{}                           &                      &                         & \multicolumn{1}{r|}{}                           &                      &                         &                             \\
\multicolumn{1}{l|}{} &  & \multicolumn{1}{l|}{} &  & \multicolumn{3}{l}{$original\_msg$}        &  & \multicolumn{1}{r|}{\multirow{-2}{*}{2}}   & \multicolumn{1}{l}{} & \multirow{-2}{*}{52}    & \multicolumn{1}{r|}{\multirow{-2}{*}{(26.00x)}} & \multicolumn{1}{l}{} & \multirow{-2}{*}{160}   & \multicolumn{1}{r|}{\multirow{-2}{*}{(80.00x)}} & \multicolumn{1}{l}{} & \multirow{-2}{*}{280}   & \multirow{-2}{*}{(140.00x)} \\
\multicolumn{1}{l|}{} &  & \multicolumn{5}{l}{\textbf{Hash $\mathcal{G}$ (SHA3-512)}}            &  & \multicolumn{1}{r|}{13}                    &                      & 123                     & \multicolumn{1}{r|}{(9.46x)}                    &                      & 242                     & \multicolumn{1}{r|}{(18.61x)}                   &                      & 379                     & (29.15x)                    \\
\multicolumn{1}{l|}{} &  & \multicolumn{5}{l}{\textbf{CPA-PKE-Encryption}}                       &  & \multicolumn{1}{r|}{764}                   &                      & 1,903                   & \multicolumn{1}{r|}{(2.49x)}                    &                      & 3,482                   & \multicolumn{1}{r|}{(4.55x)}                    &                      & 5,241                   & (6.85x)                     \\
\multicolumn{1}{l|}{} &  & \multicolumn{1}{l|}{} &  & \multicolumn{3}{l}{Secret generation}      &  & \multicolumn{1}{r|}{29}                    &                      & 427                     & \multicolumn{1}{r|}{(14.72x)}                   &                      & 984                     & \multicolumn{1}{r|}{(33.93x)}                   &                      & 1,666                   & (57.44x)                    \\
\multicolumn{1}{l|}{} &  & \multicolumn{1}{l|}{} &  & \multicolumn{1}{l|}{} &  & XOF (SHAKE-128) &  & \multicolumn{1}{r|}{25}                    &                      & 245                     & \multicolumn{1}{r|}{(9.80x)}                    &                      & 484                     & \multicolumn{1}{r|}{(19.36x)}                   &                      & 756                     & (30.24x)                    \\
\multicolumn{1}{l|}{} &  & \multicolumn{1}{l|}{} &  & \multicolumn{1}{l|}{} &  & CBD ($\beta_2$) &  & \multicolumn{1}{r|}{4}                     &                      & 182                     & \multicolumn{1}{r|}{(45.50x)}                   &                      & 499                     & \multicolumn{1}{r|}{(124.75x)}                  &                      & 909                     & (227.25x)                   \\
\multicolumn{1}{l|}{} &  & \multicolumn{1}{l|}{} &  & \multicolumn{3}{l}{Polynomial arithmetic}  &  & \multicolumn{1}{r|}{}                      &                      &                         & \multicolumn{1}{r|}{}                           &                      &                         & \multicolumn{1}{r|}{}                           &                      &                         &                             \\
\multicolumn{1}{l|}{} &  & \multicolumn{1}{l|}{} &  & \multicolumn{3}{l}{$arrange\_msg$}         &  & \multicolumn{1}{r|}{}                      & \multicolumn{1}{l}{} & \multirow{-2}{*}{1,187} & \multicolumn{1}{r|}{}                           & \multicolumn{1}{l}{} & \multirow{-2}{*}{1,640} & \multicolumn{1}{r|}{}                           & \multicolumn{1}{l}{} & \multirow{-2}{*}{2,086} &                             \\
\multicolumn{1}{l|}{} &  & \multicolumn{1}{l|}{} &  & \multicolumn{3}{l}{Polynomial Comparison}  &  & \multicolumn{1}{r|}{\multirow{-3}{*}{734}} & \multicolumn{1}{c}{} & 287                     & \multicolumn{1}{r|}{\multirow{-3}{*}{(2.00x)}}  & \multicolumn{1}{l}{} & 856                     & \multicolumn{1}{r|}{\multirow{-3}{*}{(3.40x)}}  & \multicolumn{1}{l}{} & 1,488                   & \multirow{-3}{*}{(4.86x)}   \\
\multicolumn{1}{l|}{} &  & \multicolumn{5}{l}{Other operations}                                  &  & \multicolumn{1}{r|}{112}                   &                      & 113                     & \multicolumn{1}{r|}{(1.00x)}                    &                      & 113                     & \multicolumn{1}{r|}{(1.00x)}                    &                      & 113                     & (1.00x)                     \\ \hline
\end{tabular}%
}

\end{table}

\begin{table}[!ht]
\centering
\caption{Random number requirement for all the masked schemes of Scabbard}
\label{tab:Scabbard_random}
\resizebox{\textwidth}{!}{%
\begin{tabular}{llllllllrrrlrrlrrlrrlrrlrllrrlrrlrl}
\hline
                      &  &                       &  &                       &  &                        &  & \multicolumn{26}{c}{\# Random bytes}                                                                                                                                                                                                                                                                                                                                                                                                                                                                                                                                                                                                                   &                      \\ \cline{9-35} 
                      &  &                       &  &                       &  &                        &  & \multicolumn{8}{c}{{\color[HTML]{000000} \textbf{Florete}}}                                                                                                                                    & \multicolumn{1}{l|}{{\color[HTML]{000000} }} & \multicolumn{8}{c}{{\color[HTML]{000000} \textbf{Espada}}}                                                                                                                 & \multicolumn{1}{l|}{{\color[HTML]{000000} }} & \multicolumn{8}{c}{{\color[HTML]{000000} \textbf{Sable}}}                                                                                                                  &                      \\ \cline{9-35} 
                      &  & \multicolumn{5}{c}{Order}                                                    &  & \multicolumn{1}{l}{} & \multicolumn{1}{c}{1st} & \multicolumn{1}{c}{} &                      & \multicolumn{1}{c}{2nd} & \multicolumn{1}{c}{} &                      & \multicolumn{1}{c}{3rd} & \multicolumn{1}{l|}{}                        &  & \multicolumn{1}{c}{1st} & \multicolumn{1}{c}{} & \multicolumn{1}{c}{} & \multicolumn{1}{l}{2nd} & \multicolumn{1}{c}{} & \multicolumn{1}{c}{} & \multicolumn{1}{c}{3rd} & \multicolumn{1}{l|}{}                        &  & \multicolumn{1}{c}{1st} & \multicolumn{1}{c}{} & \multicolumn{1}{c}{} & \multicolumn{1}{l}{2nd} & \multicolumn{1}{l}{} & \multicolumn{1}{c}{} & \multicolumn{1}{c}{3rd} & \multicolumn{1}{c}{} \\ \hline
\multicolumn{7}{l}{{\color[HTML]{000000} \textbf{CCA-KEM-Decapsulation}}}                               &  &                      & 15,824                  &                      &                      & 52,176                  &                      &                      & 101,280                 & \multicolumn{1}{r|}{}                        &  & 11,496                  &                      &                      & 39,320                  &                      &                      & 85,296                  & \multicolumn{1}{l|}{}                        &  & 12,496                  &                      &                      & 39,152                  &                      &                      & 75,232                  &                      \\
\multicolumn{1}{l|}{} &  & \multicolumn{5}{l}{\textbf{CPA-PKE-Decryption}}                              &  &                      & 2,560                   &                      &                      & 10,176                  &                      &                      & 20,352                  & \multicolumn{1}{r|}{}                        &  & 304                     &                      &                      & 1,216                   &                      &                      & 2,432                   & \multicolumn{1}{l|}{}                        &  & 832                     &                      &                      & 3,328                   &                      &                      & 6,656                   &                      \\
\multicolumn{1}{l|}{} &  & \multicolumn{1}{l|}{} &  & \multicolumn{3}{l}{Polynomial arithmetic}         &  &                      & 0                       &                      &                      & 0                       &                      &                      & 0                       & \multicolumn{1}{r|}{}                        &  & 0                       &                      &                      & 0                       &                      &                      & 0                       & \multicolumn{1}{l|}{}                        &  & 0                       &                      &                      & 0                       &                      &                      & 0                       &                      \\
\multicolumn{1}{l|}{} &  & \multicolumn{1}{l|}{} &  & \multicolumn{3}{l}{Compression}                   &  &                      & 2,496                   &                      &                      & 9,984                   &                      &                      & 19,968                  & \multicolumn{1}{r|}{}                        &  & 304                     &                      &                      & 1,216                   &                      &                      & 2,432                   & \multicolumn{1}{l|}{}                        &  & 832                     &                      &                      & 3,328                   &                      &                      & 1,152                   &                      \\
\multicolumn{1}{l|}{} &  & \multicolumn{1}{l|}{} &  & \multicolumn{3}{l}{$original\_msg$}               &  & \multicolumn{1}{l}{} & 64                      &                      & \multicolumn{1}{r}{} & 192                     &                      & \multicolumn{1}{r}{} & 384                     & \multicolumn{1}{l|}{}                        &  & 0                       &                      & \multicolumn{1}{r}{} & 0                       &                      & \multicolumn{1}{r}{} & 0                       & \multicolumn{1}{l|}{}                        &  & 0                       &                      & \multicolumn{1}{r}{} & 0                       &                      & \multicolumn{1}{r}{} & 0                       &                      \\
\multicolumn{1}{l|}{} &  & \multicolumn{5}{l}{\textbf{Hash $\mathcal{G}$ (SHA3-512)}}                   &  &                      & 192                     &                      &                      & 576                     &                      &                      & 1,152                   & \multicolumn{1}{r|}{}                        &  & 192                     &                      &                      & 576                     &                      &                      & 1,152                   & \multicolumn{1}{l|}{}                        &  & 192                     &                      &                      & 576                     &                      &                      & 67,424                  &                      \\
\multicolumn{1}{l|}{} &  & \multicolumn{5}{l}{\textbf{CPA-PKE-Encryption}}                              &  &                      & 13,072                  &                      &                      & 41,424                  &                      &                      & 79,776                  & \multicolumn{1}{r|}{}                        &  & 11,000                  &                      &                      & 37,528                  &                      &                      & 81,712                  & \multicolumn{1}{l|}{}                        &  & 11,472                  &                      &                      & 35,248                  &                      &                      & 6,656                   &                      \\
\multicolumn{1}{l|}{} &  & \multicolumn{1}{l|}{} &  & \multicolumn{3}{l}{Secret generation}             &  &                      & 6,528                   &                      &                      & 16,512                  &                      &                      & 29,952                  & \multicolumn{1}{r|}{}                        &  & 4,896                   &                      &                      & 14,688                  &                      &                      & 35,520                  & \multicolumn{1}{l|}{}                        &  & 6,528                   &                      &                      & 16,512                  &                      &                      & 29,952                  &                      \\
\multicolumn{1}{l|}{} &  & \multicolumn{1}{l|}{} &  & \multicolumn{1}{l|}{} &  & XOF (SHAKE-128)        &  &                      & 384                     &                      &                      & 1,152                   &                      &                      & 2,304                   & \multicolumn{1}{r|}{}                        &  & 768                     &                      &                      & 2,304                   &                      &                      & 4,608                   & \multicolumn{1}{l|}{}                        &  & 384                     &                      &                      & 1,152                   &                      &                      & 2,304                   &                      \\
\multicolumn{1}{l|}{} &  & \multicolumn{1}{l|}{} &  & \multicolumn{1}{l|}{} &  & CBD (Binomial Sampler) &  &                      & 6,144                   &                      &                      & 15,360                  &                      &                      & 27,648                  & \multicolumn{1}{r|}{}                        &  & 4,128                   &                      &                      & 12,384                  &                      &                      & 30,912                  & \multicolumn{1}{l|}{}                        &  & 6,144                   &                      &                      & 15,360                  &                      &                      & 27,648                  &                      \\
\multicolumn{1}{l|}{} &  & \multicolumn{1}{l|}{} &  & \multicolumn{3}{l}{Polynomial arithmetic}         &  &                      & 0                       &                      &                      & 0                       &                      &                      & 0                       & \multicolumn{1}{r|}{}                        &  & 0                       &                      &                      & 0                       &                      &                      & 0                       & \multicolumn{1}{l|}{}                        &  & 0                       &                      &                      & 0                       &                      &                      & 0                       &                      \\
\multicolumn{1}{l|}{} &  & \multicolumn{1}{l|}{} &  & \multicolumn{3}{l}{$arrange\_msg$}                &  & \multicolumn{1}{l}{} & 0                       &                      & \multicolumn{1}{r}{} & 0                       &                      & \multicolumn{1}{r}{} & 0                       & \multicolumn{1}{l|}{}                        &  & 256                     &                      & \multicolumn{1}{r}{} & 768                     &                      & \multicolumn{1}{r}{} & 2,048                   & \multicolumn{1}{l|}{}                        &  & 0                       &                      & \multicolumn{1}{r}{} & 0                       &                      & \multicolumn{1}{r}{} & 0                       &                      \\
\multicolumn{1}{l|}{} &  & \multicolumn{1}{l|}{} &  & \multicolumn{3}{l}{Polynomial Comparison}         &  &                      & 6,544                   &                      &                      & 24,912                  &                      &                      & 49,824                  & \multicolumn{1}{r|}{}                        &  & 5,848                   &                      &                      & 22,072                  &                      &                      & 44,144                  & \multicolumn{1}{l|}{}                        &  & 4,944                   &                      &                      & 18,736                  &                      &                      & 37,472                  &                      \\
\multicolumn{1}{l|}{} &  & \multicolumn{5}{l}{Other operations}                                         &  &                      & 0                       &                      &                      & 0                       &                      &                      & 0                       & \multicolumn{1}{r|}{}                        &  & 0                       &                      &                      & 0                       &                      &                      & 0                       & \multicolumn{1}{l|}{}                        &  & 0                       &                      &                      & 0                       &                      &                      & 0                       &                      \\ \hline
\end{tabular}%
}
\end{table}

\subsection{Performance Comparison of Masked Scabbard Suite with the State-of-the-Art}

We analyze the performance and random number requirements for masked decapsulation algorithms of Scabbard's schemes in comparison to the state-of-the-art masked implementations of LBC. We compare our masked Scabbard implementation with Bronchain et al.'s \cite{BronchainC22} and Bos et al.'s \cite{ho-mask-comparator-kyber-bos-2021} masked implementations of Kyber and Kundu et al.'s \cite{HO_mask_Saber} masked implementations of Saber in Table \ref{tab:masked_state-of-the-art_performance}. 

\begin{table}[!hbt]
\centering
\caption{Performance comparison of masked Scabbard implementations with the state-of-the-art}
\label{tab:masked_state-of-the-art_performance}
\resizebox{0.8\columnwidth}{!}{
\begin{tabular}{clrrrr|crr}
\hline
\multicolumn{2}{c}{{ }}                               & \multicolumn{4}{c|}{{ Performance}}                                                                                                                                         & \multicolumn{3}{c}{{ \# Randm numbers}}                                                                  \\ \cline{3-9} 
\multicolumn{2}{c}{{ }}                               & \multicolumn{4}{c|}{(x1000 clock cycles)}                                                                                                                                                       & \multicolumn{3}{c}{{ (bytes)}}                                                                           \\ \cline{3-9} 
\multicolumn{2}{c}{\multirow{-3}{*}{{ Scheme}}}       & \multicolumn{1}{c}{{ }} & \multicolumn{1}{c}{{ 1st}} & \multicolumn{1}{c}{{ 2nd}} & \multicolumn{1}{c|}{{ 3rd}} & { 1st} & \multicolumn{1}{c}{{ 2nd}} & \multicolumn{1}{c}{{ 3rd}} \\ \hline
{ Florete} & (this work)                              & { }                     & 2,621                                          & 4,844                                          & 7,395                                           & \multicolumn{1}{r}{15,824} & 52,176                                         & 101,280                                        \\
{ Espada}  & (this work)                              & { }                     & 4,335                                          & 6,838                                          & 9,861                                           & \multicolumn{1}{r}{11,496} & 39,320                                         & 85,296                                         \\
{ Sable}   & (this work)                              & { }                     & \textbf{2,431}                                 & \textbf{4,348}                                 & \textbf{6,480}                                  & \multicolumn{1}{r}{12,496} & 39,152                                         & 75,232                                         \\
{ Saber}   & \cite{HO_mask_Saber}                     & { }                     & 3,022                                          & 5,567                                          & 8,649                                           & \multicolumn{1}{r}{12,752} & 43,760                                         & 93,664                                         \\
{ uSaber}  & \cite{HO_mask_Saber}                     & { }                     & 2,473                                          & 4,452                                          & 6,947                                           & \multicolumn{1}{r}{10,544} & 36,848                                         & 79,840                                         \\
{ Kyber}   & \cite{BronchainC22}                      & \multicolumn{1}{l}{}                        & 10,018                                         & 16,747                                         & 24,709                                          & -                          & \multicolumn{1}{c}{-}                          & \multicolumn{1}{c}{-}                          \\
{ Kyber}   & \cite{ho-mask-comparator-kyber-bos-2021} & { }                     & $3,116^*$                                          & 44,347                                         & 115,481                                         & $12,072^*$                      & \multicolumn{1}{c}{902,126}                    & \multicolumn{1}{c}{2,434,170}                  \\ \hline
\end{tabular}
}
\text{*: optimized specially for the first-order masking}
\end{table}

First-, second- and third-order masked decapsulation implementations of Florete are respectively $73\%$, $71\%$, and $70\%$ faster than Bronchain et al.'s \cite{BronchainC22} masked implementation of Kyber. Bos et al. optimized their algorithm specifically for the first-order masking of Kyber. Even though it is $15\%$ slower than the first-order masked decapsulation of Florete. Bos et al.'s \cite{ho-mask-comparator-kyber-bos-2021} second- and third-order masked implementations of Kyber are respectively $89\%$ and $93\%$ slower than Florete. The random byte requirements in the masked version of Florete compared to Kyber are $94\%$ less for the second order and $95\%$ less for the third order. Florete also performs better than Saber. 
%Florete needs $13\%$ fewer clock cycles for first-order masking, $12\%$ fewer clock cycles second-order masking, and $14\%$ less clock cycles third-order masking than Saber. 
Florete needs $13\%$, $12\%$, and $14\%$ fewer clock cycles than Saber for first-, second-, and third-order masking.

Masked decapsulation implementation of Espada performs $56\%$, $59\%$, and $60\%$ better than Bronchain et al.'s \cite{BronchainC22} masked implementation of Kyber for first-, second-, and third-order, respectively. Second-, and third-order masked implementations of Espada are faster than Bos et al.'s \cite{ho-mask-comparator-kyber-bos-2021} masked Kyber by $84\%$ and $91\%$, respectively. The random bytes requirements in Espada compared to Kyber are $95\%$ less for the second-order and $96\%$ less for the third-order masking. Espada also uses fewer random numbers than Saber. Espada requires $9\%$ fewer random bytes in first-order masking, $10\%$ fewer random bytes in second-order masking, and $8\%$ fewer random bytes in third-order masking than Saber.    

We show that the masked implementation of Sable performs better than masked Kyber and Saber for first-, second-, and third-order (like Florete). Sable performs $75\%$, $74\%$, and $73\%$ better than Bronchain et al.'s \cite{BronchainC22} masked implementation of Kyber and $21\%$, $90\%$, and $94\%$ better than Bos et al.'s \cite{ho-mask-comparator-kyber-bos-2021} masked implementation of Kyber first-, second-, and third-order, respectively.  
Compared to Kyber, Sable requires $95\%$ and $96\%$ less random bytes for second- and third-order masking.
The performance of masked Sable is better than masked Saber by $19\%$ for first-order, $21\%$ for second-order, and $25\%$ for third-order masking. Masked Sable uses $2\%$, $10\%$, and $19\%$ less number of random bytes for first-, second-, and third-order than masked Saber, respectively. uSaber is a masking-friendly variant of Saber proposed during the third round of NIST submission. We notice that masked Sable is also faster than masked uSaber for arbitrary order. Masked Sable is $1\%$ faster for first-order, $2\%$ for second-order, and $6\%$ for third-order than masked uSaber. Although first- and second-order masked Sable needs more random bytes than uSaber, third-order masked Sable requires $5\%$ less random bytes than uSaber.  

Implementations of masked Scabbard schemes achieve better performance and use fewer random bytes than masked Kyber because the schemes of Scabbard use the RLWR/ MLWR problem as an underlying hard problem and Kyber uses the MLWE problem as the hard problem. The decapsulation operation of RLWR/ MLWR-based KEM has fewer components compared to the decapsulation operation of RLWE/ MLWE-based KEM due to the requirement of sampling error vectors and polynomials generations in the re-encryption step of RLWE/ MLWE-based KEMs. RLWR/ MLWR-based KEMs also benefit due to the use of power-of-two moduli. Computationally expensive components, such as A2B or B2A conversions, are cheaper when using power-of-two moduli. The schemes of Scabbard also use slightly smaller parameters than Kyber, which also contributes to achieving better performance and requirements of fewer random bytes for masked implementation of Scabbard's KEMs compared to Kyber. 

\section{Conclusions}

In this work, we presented the impact of different design decisions of LBC on masking. We analyzed each component where masking is needed and discussed each design decision's positive and negative impact on performance. 
As we mentioned at the beginning of the paper, it is possible to improve different practical aspects, such as masking overheads, by modifying the existing designs of PQC. 
This highlights the necessity of further research efforts to improve existing PQC designs.   

\noindent\textbf{Acknowledgements.} This work was partially supported by Horizon 2020 ERC Advanced Grant (101020005 Belfort), CyberSecurity Research Flanders with reference number VR20192203, BE QCI: Belgian-QCI (3E230370) (see beqci.eu), and Intel Corporation. 
Angshuman Karmakar is funded by FWO (Research Foundation – Flanders) as a junior post-doctoral fellow (contract number 203056 / 1241722N LV).

\bibliographystyle{splncs04}
% \bibliography{mask,crypto/crypto}
\bibliography{main}

\end{document}